\documentclass[aps,twocolumn,floatfix,altaffilletter,superscriptaddress,preprintnumbers,tightenlines,showpacs,showkeys,notitlepage,nofootinbib]{revtex4-1}

\usepackage[T1]{fontenc}
\usepackage[colorlinks=true,citecolor=blue,linkcolor=blue,urlcolor=blue]{hyperref}
\usepackage{braket,graphicx,physics,amsthm,amssymb,amsfonts,xcolor,booktabs,siunitx,caption,subcaption,multirow,tablefootnote}
\usepackage{mathtools}
\usepackage{slashed,bm}
\usepackage[capitalise, english]{cleveref}
\usepackage[normalem]{ulem}
\usepackage{xspace}
\usepackage{mathrsfs}
\usepackage{soul}
\usepackage{subcaption}
\usepackage{fontawesome}
\newcolumntype{C}[1]{>{\centering\let\newline\\\arraybackslash\hspace{0pt}}m{#1}}

\usepackage{tikz}
\usetikzlibrary{arrows,shapes}
\usetikzlibrary{trees}
\usetikzlibrary{snakes}
\usetikzlibrary{matrix,arrows} 													
\usetikzlibrary{positioning}				
\usetikzlibrary{calc,through}				
\usetikzlibrary{decorations.pathreplacing} 
\usepackage[tikz]{bclogo} 					
\usepackage{pgffor}							
\usetikzlibrary{decorations.pathmorphing}	
\usetikzlibrary{decorations.markings}
\usetikzlibrary{intersections}		

\tikzset{
    vector/.style={decorate, decoration={snake}, draw},
    fermion/.style={postaction={decorate},
        decoration={markings,mark=at position .55 with {\arrow{>}}}},
    fermionbar/.style={draw, postaction={decorate},
        decoration={markings,mark=at position .55 with {\arrow{<}}}},
    fermionnoarrow/.style={},
    gluon/.style={decorate,
        decoration={coil,amplitude=4pt, segment length=5pt}},
    scalar/.style={dashed, postaction={decorate},
        decoration={markings,mark=at position .55 with {\arrow{>}}}},
    scalarbar/.style={dashed, postaction={decorate},
        decoration={markings,mark=at position .55 with {\arrow{<}}}},
    scalarnoarrow/.style={dashed,draw},
	vectorscalar/.style={loosely dotted,draw=black, postaction={decorate}},
}

\makeatletter
\providecommand*{\diff}%
	{\@ifnextchar^{\DIfF}{\DIfF^{}}}
\def\DIfF^#1{%
	\mathop{\mathrm{\mathstrut d}}%
		\nolimits^{#1}\gobblespace}
\def\gobblespace{%
	\futurelet\diffarg\opspace}
\def\opspace{%
	\let\DiffSpace\!%
	\ifx\diffarg(%
		\let\DiffSpace\relax
	\else
		\ifx\diffarg[%
			\let\DiffSpace\relax
		\else
  			\ifx\diffarg\{%
				\let\DiffSpace\relax
			\fi\fi\fi\DiffSpace}

\definecolor{cbred}{HTML}{ff0000}
\definecolor{cborange}{HTML}{FFA500}
\definecolor{cbgreen}{HTML}{008000}
\definecolor{cbyellow}{HTML}{f1dd42}
\definecolor{cblblue}{HTML}{56b4e9}
\definecolor{cbblue}{HTML}{0072b2}
\definecolor{defgrey}{HTML}{9f9f9f}
\definecolor{defgreen}{HTML}{8eba42}
\definecolor{defgrey}{HTML}{808080}
\definecolor{ercolor}{HTML}{00FFFF}
\definecolor{nrcolor}{HTML}{FF00FF}
\definecolor{sigcolor}{HTML}{FFE600}
\definecolor{yellow35}{HTML}{ffae17}
\definecolor{green35}{HTML}{9fac43}
\definecolor{blueWIMP}{HTML}{0000aa}

\definecolor{Tsallis}{HTML}{dfa8be}
\definecolor{Empirical}{HTML}{f1cb98}
\definecolor{Boltzmann}{HTML}{9ccff1}

\definecolor{fig34red}{HTML}{99434b}
\definecolor{fig34blue}{HTML}{4169E1}

\newcommand{\Leff}{\Lambda_{\text{UV}}}
\newcommand{\mdm}{m_\text{DM}}
\newcommand{\hc}{\text{h.c.}}
\newcommand{\chidm}{\chi_\text{DM}}
\newcommand{\vrel}{v_\text{rel}}

\newcommand{\rvline}{\hspace*{-\arraycolsep}\vline\hspace*{-\arraycolsep}}
 
\def\l@subsection#1#2{}
\def\l@subsubsection#1#2{}

\begin{document}


\title{Looking for WIMPs through the neutrino fogs}

\author{Itay M. Bloch}
\affiliation{Theory Group, Lawrence Berkeley National Laboratory, Berkeley, CA 94720, U.S.A.\\ Berkeley Center for Theoretical Physics, University of California, Berkeley, CA 94720, U.S.A.}
\author{Salvatore Bottaro}
\affiliation{Raymond and Beverly Sackler School of Physics and Astronomy, Tel-Aviv 69978, Israel}
\author{Diego Redigolo}
\affiliation{INFN, Sezione di Firenze Via G. Sansone 1, 50019 Sesto Fiorentino, Italy}
\author{Ludovico Vittorio}
\affiliation{LAPTh, Université Savoie Mont-Blanc et CNRS, 74941 Annecy, France}

\begin{abstract}
We revisit the expected sensitivity of large-scale xenon detectors to Weakly Interacting Massive Particles (WIMPs). Assuming current primary noise sources can be mitigated, we find that with the present discrimination power between nuclear and electron recoils, the experimental sensitivity is limited not only by atmospheric neutrinos' nuclear recoils (``nuclear recoil neutrino fog'') but also by solar neutrinos' electron-recoil events (``electron recoil neutrino fog''). While this is known by experimentalists, it is often missed or misunderstood by theorists, and we therefore emphasize this effect. We set up a realistic detector simulation to quantify the contamination of the WIMP signal from both these neutrino backgrounds. We observe that the electron-recoil background remains significant even for signal rates exceeding those of atmospheric neutrinos, as predicted by most electroweak WIMP candidates. We update the projections for the required exposure to exclude/discover a given electroweak WIMP, streamlining the computation of their signal rates and uncertainties. We show that all of the real WIMPs with zero hypercharge can be excluded (discovered) with a 50 tonne year (300 tonne year) exposure. A similar exposure will allow to probe a large portion of the viable parameter space for complex WIMP with non-zero hypercharge.
\end{abstract}


\maketitle


\section{Introduction}

The WIMP paradigm is one of the most appealing and predictive theoretical frameworks to explain the existence of Dark Matter (DM). In this setup, the DM is in thermal equilibrium in the early Universe and freezes out purely through annihilations into SM states. In the simplest realization of this paradigm, the DM is part of a single electroweak (EW) multiplet and the requirement of reproducing the observed dark matter abundance today gives a precise prediction for the dark matter mass. Considering different EW multiplets up to the perturbative unitarity bound~\cite{Griest:1989wd}, the resulting mass range spanned by EW WIMPs lies between $1$ TeV and $300$ TeV as summarized in Ref.~\cite{Bottaro:2021snn,Bottaro:2022one}.   

It is then natural to ask whether or not we will be able to make a conclusive statement about the existence of EW WIMPs and on which time scale. In this work, we focus on present and future direct detection experiments looking for dark matter recoil onto xenon. Our goal is to revisit the projected sensitivity of the next generation of large-scale xenon experiments~\cite{Mount:2017qzi,XENON:2020kmp,DARWIN:2016hyl,Aalbers:2022dzr}. 

For heavy DM masses, it is sometimes assumed in theory projections that the DM signal is background-free until its rate becomes comparable to the \emph{irreducible} background given by the coherent elastic scattering of neutrinos onto nuclei. This background is dominated by neutrinos produced by cosmic rays scattering onto the atmosphere and by diffuse supernova emission which together form the so-called neutrino floor~\cite{Monroe:2007xp,Vergados:2008jp,Strigari:2009bq,Gutlein:2010tq}, more recently dubbed ``neutrino fog''~\cite{OHare:2021utq}. For the rest of the paper, we hereby term this the ``nuclear recoil neutrino fog''. 

Here we go beyond the standard assumption, including in our projections the background from solar neutrino electron recoils which leaks into the WIMP signal region because of the finite experimental discrimination between nuclear recoils and electron recoils~\cite{Billard:2013qya,Newstead:2020fie}. Even if the electron recoil background is \emph{reducible}, its rate is so much larger than the standard nuclear recoil neutrino fog that it cannot be neglected in the assessment of the expected reach. Here we improve upon the analysis of Ref.~\cite{Gaspert:2021gyj} making use of the realistic simulation environment based on the Noble Element Simulation Technique (NEST) package~\cite{2011arXiv1106.1613S,szydagis_2018_1314669}\footnote{We use the python package nestpy (v2.0.2), which runs the NEST (v2.4.0) code. See \href{https://github.com/ItayBM/NeutrinoFogs}{github} for the full analysis details.}. We compute the leakage of electron recoil events into nuclear recoil and perform a two-dimensional likelihood analysis in the scintillation signals observed in xenon detectors to derive the required exposure to exclude or discover a given signal rate. As expected, we find that our analysis outperforms a simple cut-and-count approach, though it also outperforms an ER-less analysis with a $50\%$ acceptance efficiency in some cases (a procedure often used by theorists for low signal rates~\cite{DARWIN:2016hyl,Aalbers:2022dzr}). Our result is subject to changes when experimental systematics are taken into account.  We base all our simulations on an LZ-like detector~\cite{LZ:2022lsv}. While future detectors may offer enhanced discrimination power, as noted in Ref.~\cite{Newstead:2020fie}, the mean lifetime of drifting electrons is the most important parameter. Since this lifetime already exceeds $5~{\rm msec}$, significant improvements in discrimination are unlikely without adopting more novel techniques, such as directional detection methods~\cite{OHare:2022jnx}.

\begin{figure*}[htp!]
    \centering
    \includegraphics[width=0.8\textwidth]{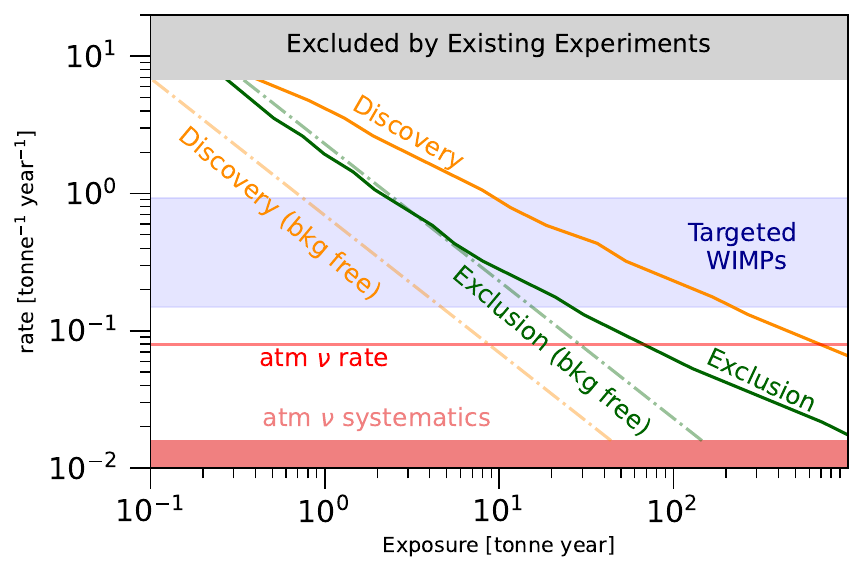}
    \caption{The signal in inverse tonne-year for a $5\sigma$ discovery ({\color{cborange}\bf in orange}) and a $90\%$ exclusion ({\color{cbgreen}\bf in green}) as a function of the required exposure. The {\bf solid lines} correspond to the realistic sensitivity derived here while the {\bf dashed lines} correspond to the naive results of a background-free analysis (see Sec.~\ref{sec:results} for the details). The slight improvement in the finite background LLR compared to the zero-background case at small exposures is a well-known artifact of the LLR method when used for projections, see App.~\ref{sec:analysisdetails} for further discussion. The {\color{defgrey}\bf grey band} on the right shows the present excluded rate from existing experiments~\cite{LZ:2022lsv,PandaX-4T:2021bab,XENON:2023cxc} with the strongest constraint currently from LZ~\cite{LZ:2022lsv}. The {\color{cbred}\bf red line} corresponds to the atmospheric neutrino rate while the {\color{cbred}\bf red band} indicates where the expected rate lies below the uncertainty in the atmospheric neutrino flux. The {\color{blueWIMP}\bf blue band} corresponds to the envelope of the different expected rates for real EW WIMPs with $\rho_{\rm{DM}}=0.3\,\rm{ GeV}/\rm{cm}^3$ is the DM local energy density.}
    \label{fig:ExclusionDiscovery}
\end{figure*}

Figure~\ref{fig:ExclusionDiscovery} shows our final result which is the signal discovered or excluded for a hypothetical large-scale xenon detector as a function of exposure.

The targeted EW WIMPs rates are shown as a blue band, where for reference we consider the predictions of real EW WIMPs with zero hypercharge: the bottom of the band is the minimal cross-section of the 3-plet while the top of the band is the maximal cross-section of the real 13-plet. As shown in the plot, an exposure of a few hundred tonne-years will be sufficient to exclude or discover all the real EW WIMP thermal candidates. Such an exposure is expected to be achieved at future large-scale detectors such as XLZD~\cite{Aalbers:2022dzr,DARWIN:2016hyl,Baudis:2024jnk} or PandaX-xT~\cite{PandaX:2024oxq}. We zoom in Fig.~\ref{fig:rate_money5and3} on the particularly theoretically motivated cases of the real 3-plet and 5-plet~\cite{Arkani-Hamed:2004ymt,Giudice:2004tc,Cirelli:2005uq} showing the difference with the background free projections. 

We further explore the implications of our result in the parameter space of \emph{all} the EW WIMPs, including the complex WIMPs with non-zero hypercharge. The required exposure to probe all the SU(2) $n$-plet is shown in Fig.~\ref{fig:rate_money2}, where for complex WIMPs we fix the couplings with the Higgs to their minimal allowed value~\cite{Bottaro:2022one}. Fig.~\ref{fig:mass_splitting} summarizes instead the status of complex EW WIMPs for all the allowed range of couplings with the SM Higgs. In particular, we find that tuning the cross section below the neutrino floor is always possible for complex EW WIMPs besides for the 10-plet and the 12-plet with $Y=1/2$ and for the 5-plet with $Y=1$. 

As summarized in Fig.~\ref{fig:deltaR}, we assess the uncertainties in the WIMP signal rate including the theory uncertainties on the freeze-out prediction of the DM mass, the uncertainties on the WIMP cross section on nuclei and the ones originating from our ignorance about the DM velocity distribution. Throughout the paper we fix the DM local energy density to $\rho_{\rm{DM}}=0.3\,\rm{GeV}/\rm{cm}^3$~\cite{Baxter:2021pqo} but derive simple scalings of our results with $\rho_{\rm{DM}}$ in order to make it possible to account for the present large uncertainties on this parameter. 

As shown in Fig.~\ref{fig:sigmaplot}, we update the computation and the theory uncertainty on the EW WIMP cross section with nuclei for all the EW multiplets. We show that the uncertainty originating from the theory errors on the lattice nuclear form factors can be of the same order of the astrophysical uncertainty on the DM velocity distribution\footnote{As detailed in Sec.~\ref{sec:astro} the astrophysical uncertainties on the velocity distribution adopted here are sensibly larger compared to the ones suggested in the 
recommended conventions of Ref.~\cite{Baxter:2021pqo}.} and can even exceed it if the current discrepancy among the quark nuclear form factors computed in different flavor schemes is taken as a theoretical systematic.  

This paper is organized as follows: after reviewing the basics of how to map any differential rate in energy to a scintillation signal in xenon we discuss the neutrino backgrounds and the DM signal and describe our main results in Sec.~\ref{sec:setup}. In Sec.~\ref{sec:Signal} we map this result in the EW WIMP parameter space and summarize the basics of WIMP signal rate and uncertainties in Sec.~\ref{sec:Signalerror}. In the Appendix~\ref{sec:analysisdetails} we compare our analysis to simpler alternatives. A running python notebook with our analysis setup, as well as a mathematica file with sufficient information to reproduce our analysis can be found in \href{https://github.com/ItayBM/NeutrinoFogs}{github}. In order to make the paper self-contained, the Appendix~ \ref{app:theory} summarizes many technical details on the theoretical computation of the signal rate.

\section{Signal, Backgrounds and Analysis}\label{sec:setup}

 We summarize the basics of the theory connecting a given rate of collisions with xenon nuclei or electrons in xenon to the signal detected in the photomultiplier tubes (PMTs) in large-volume xenon detectors. The detectors currently used by the leading direct detection experiments use a technology known as dual-phase Time Projection Chambers (TPCs) where most of the detector mass is in a liquid phase while a part of it is in the gas phase~\cite{XENON:2010xwm,XENON100:2011cza,XENON:2017lvq,XENON:2024wpa}. 
 
When energy is deposited in the liquid xenon target, two signals are generated and then measured by PMTs. The first signal (primary scintillation, S1) comes from detecting the primary scintillation photons. The second signal (proportional scintillation, S2) arises from photons emitted when ionization electrons (after being drifted from the initial interaction site by an electric field) are accelerated in the gas phase.
The raw S1,S2 signals are later corrected from different detector distortions to the ``corrected S1'' and ``corrected S2" (cS1 and cS2 respectively) in a process beyond the scope of this work (see e.g.~\cite{XENON:2024qgt}). This two-dimensional signal allows the separation between electron recoil events and nuclear recoil events. 

 This section is organized as follows. We first present the signal rate from DM nuclear recoil in Sec.~\ref{sec:DMsignal} and then move to discuss the backgrounds considered here in Sec.~\ref{sec:neutrinobkd} and give a summary of the other backgrounds in Sec.~\ref{sec:extra}. Sec.~\ref{sec:results} illustrates the results of our analysis.  

\subsection{Dark Matter signal}\label{sec:DMsignal}
The differential rate of DM collisions per nuclear recoil energy, $d R_\chi/d E_R$, can be related to the expected measured density of events in the $({\rm cS1},{\rm cS2})$ channels of xenon detectors, $\mathcal{R}_\chi({\rm cS1}, {\rm cS2})$, as 
\begin{equation}
\mathcal{R}_\chi({\rm cS1}, {\rm cS2})\!= \!\text{exp} \!\!\int\!\! \frac{d R_\chi}{d E_R} \epsilon_N(E_R) \mathcal{P}_N({\rm cS1},{\rm cS2}\vert E_R) d E_R\,,\label{eq:signalrate}
\end{equation}
where the exposure of a given experiment is indicated as ``exp'' and factorized in front so that $d R_\chi/d E_R$ is the differential rate of DM scattering with the nucleus per recoil energy $E_R$ and per unit target mass. $\mathcal{P}_N({\rm cS1},{\rm cS2}\vert E_R)$ is the probability distribution associating a ${\rm cS1}$ and ${\rm cS2}$ signal for every nuclear recoil with recoil energy $E_R$ and $\epsilon_N(E_R)$ is the efficiency of the signal which ensures that the probability distribution is properly normalized. Note that if one applies cuts on the $({\rm cS1},{\rm cS2})$ space, this efficiency will be lowered if $\mathcal{P}_N({\rm cS1},{\rm cS2}|E_R)$ is normalized to 1 on the region of interest (RoI) defined by those cuts. The probability distribution and the efficiency are computed here using the NEST simulation framework~\cite{2011arXiv1106.1613S,szydagis_2018_1314669}. The data necessary to get this probability distribution are provided as an addendum to this paper in \href{https://github.com/ItayBM/NeutrinoFogs}{github}. The signal distribution in the $({\rm cS1},{\rm cS2})$ plane is shown in Fig.~\ref{fig:heatmap}.  

The differential rate per unit target mass can be written as 
\begin{equation}\label{eq:diff_rate}
\frac{d R_\chi}{d E_R}= \frac{1}{m_A}\frac{\rho_{\rm{DM}}}{m_{\rm{DM}}}\int_{v_{\text{min}}}^{v_{\text{max}}} d^3v \frac{d \sigma_\chi}{d E_R} v\tilde{f}(\vec{v},t)\ ,
\end{equation}
where $\mdm$ is the DM mass, $m_A\approx\! \langle A\rangle m_N=123.1\,\rm{GeV}$ is the mass of the xenon nucleus averaged over its isotopes, $\langle A\rangle$ the average atomic mass number and $m_N$ is the mass of the constituent nucleon (see Appendix \ref{app:composition} for the composition assumed here). $v_{\text{max}}$ is the DM escape velocity and  $v_{\text{min}}$ is the minimum velocity required to produce a recoil energy $E_R$ and depends on the kinematic of the scattering process. For elastic scattering $v_{\text{min}}=\sqrt{m_N E_R/2\mu_N^2}$ where $\mu_N$ is the reduced mass of the nucleon-DM system. The differential cross section as a function of the nuclear recoil energy $d\sigma_\chi/d E_R$ depends on the UV model. In Sec.~\ref{sec:Signalerror} we will discuss how this rate is simplified for heavy WIMPs and summarize the theoretical uncertainties on $d\sigma_\chi/dE_R$ for EW WIMPs.

$\tilde{f}(\vec v, t)$ is the velocity distribution relative to the target in the Earth frame normalized such that $\int \tilde{f}(v) d^3v=1$. This can be related to the velocity distribution in the galaxy by a Galilean boost: $\tilde{f}(\vec v,t)=f(\vec v+\vec v_\oplus(t))$, where $v_\oplus(t)$ is the relative motion of the Earth with respect to the galactic frame~\cite{Savage:2006qr}. Finally, the DM number density can be written as $\rho_{\rm{DM}}/m_{\rm{DM}}$ where $\rho_{\rm{DM}}$ is the local DM energy density. The astrophysical uncertainties considere here on both $\tilde{f}(\vec v, t)$ and $\rho_{\rm{DM}}$ are briefly summarized in Sec.~\ref{sec:astro}.

\subsection{Backgrounds from Neutrino nuclear and electron recoils}\label{sec:neutrinobkd}

Neutrinos may interact in the detector with either the nuclei or the electrons. Here we summarize these rates for completeness and refer to Refs.~\cite{Billard:2013qya,Newstead:2020fie} for more in-depth studies.  

\paragraph{Nuclear Recoils}

For neutrino nuclear recoils the same efficiency and probability distribution defined in Eq.~\eqref{eq:signalrate} can be used to map the differential rate in recoil energy to the $({\rm cS1},{\rm cS2})$ plane. The differential rate per unit target mass can be written in terms of the neutrino flux as  
\begin{equation}
\frac{d R_{\nu N}}{d E_R}=\frac{1}{m_A}\int_{E_{\nu, N}^{\rm{min}}} \frac{d \Phi_\nu}{d E_\nu} \frac{d\sigma_{\nu N}(E_\nu, E_R)}{d E_R} d E_\nu\ ,
\end{equation}
where the EW neutrino cross section with the nuclei was first computed in Ref.~\cite{Freedman:1977xn} and $d \Phi_\nu/d E_\nu$ is the differential neutrino flux. In the limit $m_N\gg E_\nu$ one can write the minimal neutrino energy to produce a fixed nuclear recoil $E_R$, which is $E_{\nu, N}^{\rm{min}}=\sqrt{m_N E_R/2}$. 

Three distinct sources of neutrinos contribute significantly to the neutrino flux: i) solar neutrinos, ii) atmospheric neutrinos, iii) the diffuse background of neutrinos from supernovae (often dubbed DSNB neutrinos). 

\begin{figure}[htp!]
    \centering
    \includegraphics[width=\linewidth]{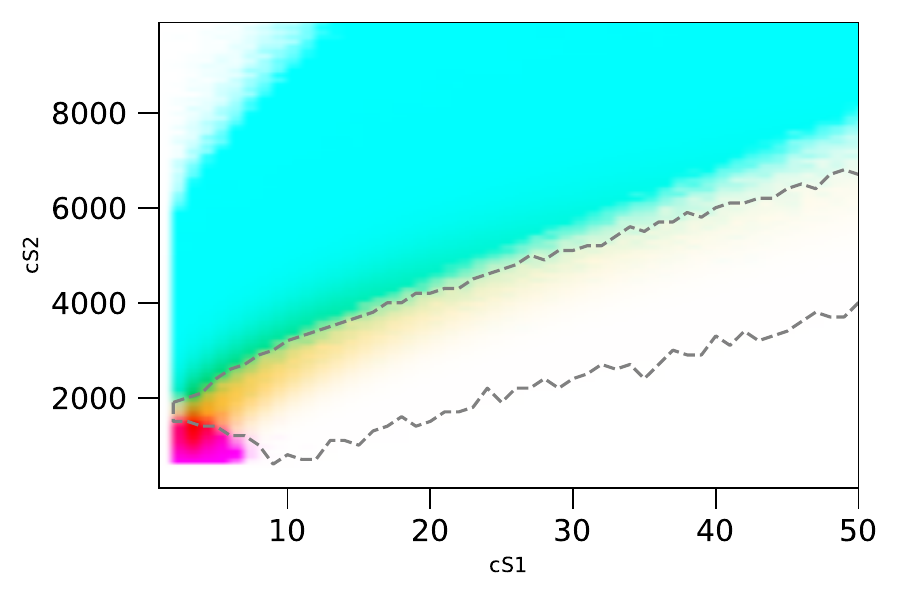}
    \caption{Distributions in the $({\rm cS1},{\rm cS2})$ plane of the heavy WIMP signal, NR Background from solar neutrinos ({\color{nrcolor}\bf in magenta}) and the ER background from solar neutrinos ({\color{ercolor}\bf in cyan}). The dashed gray line shows the optimal region of interest which maximizes the reach for a simple cut-and-count analysis. The background from atmospheric neutrinos would look identical to the signal ({\color{sigcolor}\bf in yellow}) but its rate is much below the targeted signal rates in this paper (see Fig.~\ref{fig:ExclusionDiscovery}). Note that to be able to show all contributions in a single plot, the number of background events in one bin is truncated at the maximal number of signal events per bin. See \href{https://github.com/ItayBM/NeutrinoFogs}{github} for details.}
    \label{fig:heatmap}
\end{figure}

The solar neutrinos flux is dominantly produced at energies $E_\nu\lesssim 20~{\rm MeV}$, with most NR events with recoil energy $E_R\lesssim 6~{\rm keV}$. These events are a background for light WIMPs with masses around a few GeVs. Since we consider WIMPs which are far more massive ($\mdm\gtrsim 1\, \rm{ TeV}$), cutting away the low energy end of the WIMP spectrum we may safely forget about the solar neutrino background in our analysis. This is clearly shown in Fig.~\ref{fig:heatmap} where the magenta region at low ${\rm cS1}$ and low ${\rm cS2}$ correspond to the solar neutrino NR background. 

The atmospheric neutrinos are produced when cosmic rays collide with the Earth's atmosphere, producing pions which then decay, producing neutrinos. The atmospheric neutrino flux is quite flat up to $E_\nu\sim 100~{\rm MeV}$ and starts decreasing at higher energies (see Ref.~\cite{Strigari:2009bq} for the flux). These scattering events have a spectrum very similar to WIMPs with a mass $\mdm\gtrsim 100~{\rm GeV}$ and constitute the main irreducible background for the heavy WIMPs signal considered here with an uncertainty in the flux around $20\%$~\cite{Baxter:2021pqo}. The DSNB neutrinos are a subdominant component of the background with a large $50\%$ uncertainty in their flux~\cite{Billard:2013qya}. 

The systematic uncertainties on the NR neutrino background constitute the main obstruction to reaching DM rates per unit target mass below 0.1 event per tonne per year. Luckily, the EW WIMPs rates targeted in this study are larger and these systematics can be neglected. 

\paragraph{Electron Recoils} For neutrino electron recoils, the measured rate at xenon detectors $\mathcal{R}_{\nu e}({\rm cS1}, {\rm cS2})$, can be related to the differential rate per target recoil energy $d R_{\nu e}/d E_r$ as 
\begin{equation}
\mathcal{R}_{\nu e}({\rm cS1}, {\rm cS2})\!= \!\text{exp} \!\!\int\!\! \frac{d R_{\nu e}}{d E_r} \epsilon_e(E_r) \mathcal{P}_e({\rm cS1},{\rm cS2}\vert E_r) d E_r\,,\label{eq:bkdrate}
\end{equation}
where the probability distribution of an $({\rm cS1},{\rm cS2})$ signal at a given recoil energy $E_r$ is encoded in $\mathcal{P}_e({\rm cS1},{\rm cS2}\vert E_r)$ with the efficiency $\epsilon_e(E_r)$ to provide the proper normalization. Note that if one applies a cut on the $({\rm cS1},{\rm cS2})$ space, this efficiency will be lowered if $\mathcal{P}_e({\rm cS1},{\rm cS2}|E_r)$ is normalized to 1 on the remaining RoI. The necessary data to get the probability distribution and efficiency are provided in \href{https://github.com/ItayBM/NeutrinoFogs}{github}.

The differential rate per unit target mass can be written in terms of the neutrino flux as 
\begin{equation}
\frac{d R_{\nu e}}{d E_r}=\frac{1}{m_A}\int_{E_{\nu, e}^{\rm{min}}}\frac{d \Phi_\nu}{d E_\nu} \frac{d\sigma_{\nu Z}(E_\nu,E_r)}{d E_r} d E_\nu\ ,
\end{equation}
where two different processes contribute to the neutrino-electron scattering: the neutral
current Z-exchange that is possible for all three neutrino flavors and the charged current W-exchange which requires an electron neutrino. The complete formula can be found in Ref.~\cite{Formaggio:2012cpf}, under the assumption that the bound electrons collided with may be treated as free ones. More detailed calculations implementing deviations from the free-electron limit have been performed in Ref.~\cite{Chen:2016eab} whose rates we used in this analysis.

For pedagogical reasons, let us briefly describe the calculation within the free-electron approximation, in order to get a rough understanding of the rate and kinematics of the neutrino electron recoils. In this approximation, the minimal neutrino energy for a given recoil energy is given by $E_{\nu, e}^{\rm{min}}=\left(E_r+\sqrt{E_r(E_r+2 m_e)}\right)/2$. As a consequence, the neutrino fluxes that contribute to the scattering rate in the region of interest are all solar in nature. These neutrinos have a flux which is overwhelmingly higher than the atmospheric one with energies below $\sim {\rm MeV}$ and they create ER events within a broad range of energies. For the relevant $1-15~{\rm keV}$ electron-recoil energy range, they create roughly two to three events per keV per tonne per year. 

The actual rate in the region of interest for a given DM search depends crucially on the experimental rejection which is better evaluated in the $({\rm cS1},{\rm cS2})$ plane as shown in Fig.~\ref{fig:heatmap} because of the intrinsic 2D nature of the signal. In Sec.~\ref{sec:results} we will quantify the consequences of the existence of this background on the expected sensitivity to the WIMP signal improving on a first estimate made in Ref.~\cite{Gaspert:2021gyj}. 

In our analysis we neglect the possible systematic uncertainties coming from the mapping of electron recoil events in the $({\rm cS1},{\rm cS2})$ plane. As an illustrative example, we note the recent measurement of Ref.~\cite{Temples:2021jym}, which showed that the currently used $\beta$-sources for calibrating responses to ER neutrino events, underestimate the $({\rm cS1},{\rm cS2})$ spread in certain energies. Further studies are required to ensure that this source of systematic uncertainty is small enough to not affect the projected reach on EW WIMPs. 

\subsection{Further backgrounds}\label{sec:extra}
On top of the backgrounds from neutrino fluxes discussed in the previous sections, realistic large-scale xenon detectors are affected by different sources of noise~\cite{Formaggio:2004ge}. At present experiments~\cite{LZ:2022lsv,XENON:2023cxc}, in the selected region of interest more than 90\% of the backgrounds are unrelated to the neutrino fogs discussed above. The implicit assumption we are making in this study is that these noise sources will be suppressed enough in the future to be negligible with respect to the backgrounds coming from neutrinos. In the remainder of this subsection, we comment on the different sources of noise and the existing or proposed techniques to suppress them. 

The main noise source is currently from ER events produced by Radon isotopes decaying in the liquid xenon, which much like the ER neutrino fog, leak to the NR ROI~\cite{LZ:2022lsv,XENON:2023cxc,XENON:2020kmp}. Besides the performances currently achieved by different Radon reduction techniques, the intrinsic challenge of radon purification is that Radon isotopes are constantly emanating from the detector material mixing with the liquid xenon. The use of a crystalline form of xenon has been proposed in Refs.~\cite{Kravitz:2022mby,Chen:2023llu} as a possibility to drastically reduce the Radon concentration, making it subdominant to the solar neutrino ER events.

Two extra sources of background are given by the double-beta decay of the ${}^{136}{\rm Xe}$ isotope of xenon and the NR of radiogenic neutrons sourced by different detector materials. Currently these create a background which is slightly subdominant or at most of the same order of the one induced by the solar neutrino ER events. In particular, the neutron events are already effectively vetoed by surrounding the xenon detector with an external neutron detector. However, improvements are required to reduce both these backgrounds well below that of the neutrinos. 

The last two sources of background are "surface events", and Accidental Coincidence (AC) events. Surface events occur near the surface, and have an abnormally low S2/S1 (even lower than that of NR events). AC events occur when S2-only and S1-only events are accidentally tagged as a single event.
The former can be distinguished from NR by fiducializing the volume to the center of the detector because of their spatial dependence. Furthermore, since larger detectors have a smaller surface-to-volume area, this fiducialization is expected to become more efficient for future detectors. The latter background is more relevant for small DM masses, see e.g.~\cite{XENON:2024ijk}.

\begin{figure*}[htp!]
    \centering
    \includegraphics[width=0.8\textwidth]{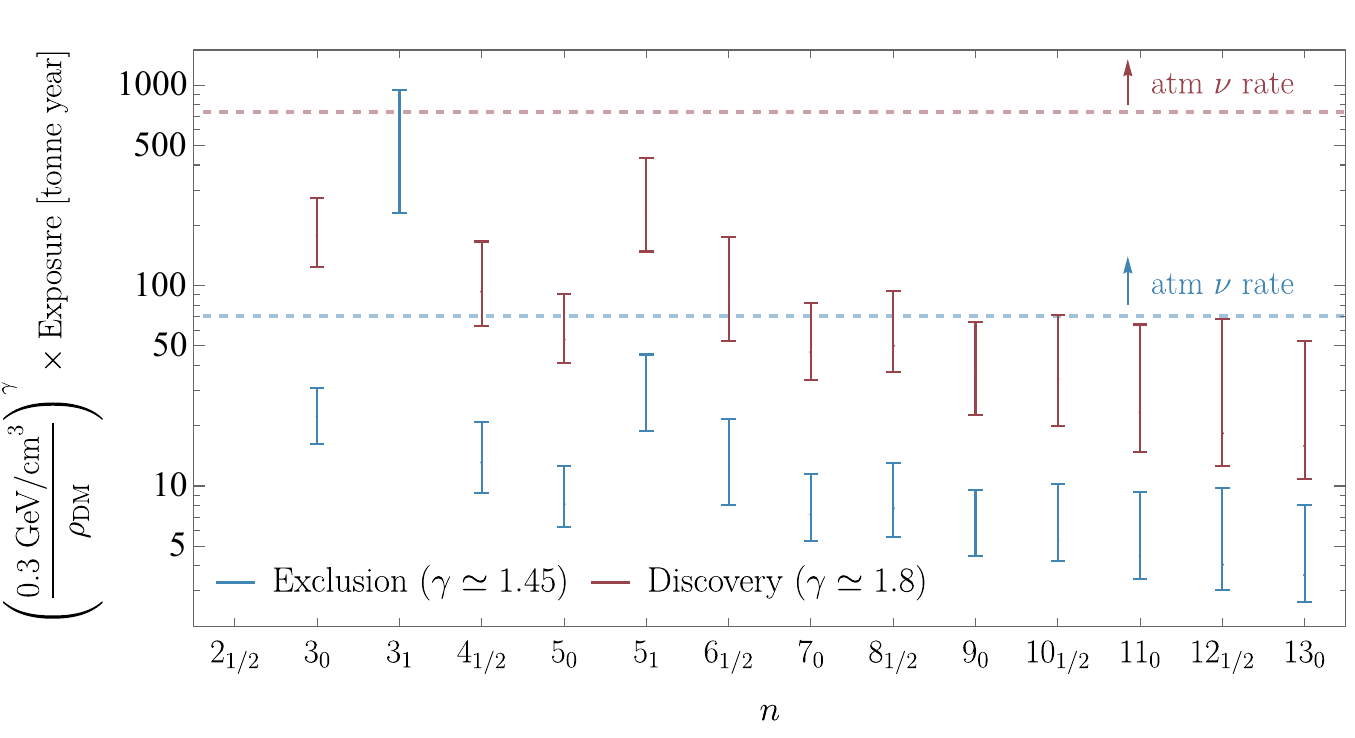}
\caption{Required exposure in tonnes year to exclude ({\color{fig34blue}\bf blue bars}) or discover ({\color{fig34red}\bf red bars}) the EW WIMPs. Whenever the required exposure exceeds the dashed line  for exclusion ({\color{fig34blue}\bf in blue}) and for discovery ({\color{fig34red}\bf in red}), the neutrino NR cannot be neglected. For the complex WIMPs with non-zero hypercharge we fix the minimal allowed splitting. The details of how this is derived are described in Sec.~\ref{sec:setup}. Our result is normalized to $\rho_{\rm{DM}}=0.3\,\rm{GeV/cm}^3$, and the scaling with exposure is derived from Fig.~\ref{fig:ExclusionDiscovery}. For each WIMP we use the prediction in the $N_f=2+1$ lattice scheme and include in our error band the mass uncertainty, the cross section uncertainty and the astrophysical uncertainties on the DM velocity distribution. Details on how these uncertainties are computed are given in Sec.~\ref{sec:Signalerror}.}
    \label{fig:rate_money2}
\end{figure*}

\subsection{Analysis}\label{sec:results}

We are now ready to discuss our analysis. We describe here the procedure to define the discovery and exclusion lines of Fig.~\ref{fig:ExclusionDiscovery}. We remind that no systematic uncertainties were accounted for in this derivation. This approximation entirely breaks down when the DM rate becomes smaller than the uncertainty in the neutrino NR rate, but accounting for systematics will have an effect much before, roughly when the systematic uncertainty in our backgrounds becomes comparable to their statistical uncertainty. As we show in Sec.~\ref{sec:neutrinobkd} most of the EW WIMPs satsify this requirement. Conversely, systematics affecting neutrino ER may require further studies to be adequately determined. 

The distribution for the backgrounds and signals in the $({\rm cS1},{\rm cS2})$ plane is shown via a heatmap in Fig.~\ref{fig:heatmap}. We also show as a black line  the boundaries of an optimal region for a simple cut and count experiment (see appendix \ref{sec:analysisdetails} for how this region was derived). NEST simulations were ran for $10^{7}$ signal-like events, solar-neutrino ER events, solar-neutrino NR events, and dsnb+atm events for a total of $4\times 10^7$ events. These simulations were then binned in the $({\rm cS1},{\rm cS2})$ space, with the size of one bin being $1\times 100$ so that the total space is tessellated by $100\times100$ bins. The resulting 2D distribution, with the amount of events in each bin assumed to be Poisson distributed with the appropriate rate. Using these distributions as the new baseline, we performed a log-likelihood ratio (LLR) test defined as  
\begin{equation}
\lambda=-2{\rm LLR}=-2\sum_{i\in{\rm bins}}\left(-s_i+n_i\log\left(1+\frac{s_i}{b_i}\right)\right),
\end{equation}
where $s_i,\,b_i,\,n_i$ are the signal,background and measured number of events in the $i$th bin respectively. 

To find the projected exclusion lines, for a given signal rate and exposure, we ran $10^4$ simulations of $\{n\}_{\rm bins}$, under the background-only and $10^4$ simulations of $\{n\}_{\rm bins}$ under the signal+background hypothesis, and found their distributions of $\lambda$s. 

The exposure for which the median of the former is equal to the $90\%$ quantile of the latter, is the one for which half of the experiments, assuming no true signal, should be able to exclude the specific signal rate. In formulas this is defined as
\begin{equation}
Q_{50\%}(\lambda|_{b,{\rm exp}_{\rm exc}})=Q_{90\%}(\lambda|_{b+s,{\rm exp}_{\rm exc}})\ ,
\end{equation}
with Q marking the quantile.

Similarly, for the discovery estimate, we estimated when the $0.000029\%$ quantile of the background-only hypothesis is equal to the median of the background+signal hypothesis, ensuring a $5\sigma$ deviation from the former half the time, assuming the latter.\footnote{Since finding the $0.000029\%$ quantile would require an unreasonable amount of simulations, we assumed the distribution of the LLR to be gaussian and estimated the location of the $0.000029\%$ quantile using its median and $5\%$ quantile: $Q_{2.9\times 10^{-7}}(\lambda|_{b,{\rm exp}_{\rm disc}}) \approx Q_{50\%}(\lambda|_{b,{\rm exp}_{\rm disc}})-3.04*(Q_{50\%}(\lambda|_{b,{\rm exp}_{\rm disc}})-Q_{5\%}(\lambda|_{b,{\rm exp}_{\rm disc}}))$. This approximation breaks down when the number of background events is too low, at which point a more accurate approximation is required, see appendix~\ref{sec:analysisdetails}.} In formula this is   
\begin{equation}
Q_{2.9\times 10^{-7}}(\lambda|_{b,{\rm exp}_{\rm disc}})=Q_{50\%}(\lambda|_{b+s,{\rm exp}_{\rm disc}})\ .    
\end{equation}

Although this constitutes an approximation, it is anticipated to deviate only marginally from a more precise estimate, the latter necessitating millions of simulations and the implementation of a comprehensive detector model incorporating the electric field profile and additional relevant parameters. Consequently, the more exact estimate was not pursued.

The dotted-dashed lines in Fig.~\ref{fig:ExclusionDiscovery} assumed zero background and can be simply written as $ s=-\log(N)$, where $s$ is the total number of signal events and $N=0.1 (0.5)$ for exclusion (discovery) corresponding to roughly 2.3 and 1 signal event. For the discovery line, this is the rate for which half of the time, at least one event will be produced by the signal (which would be sufficient for discovery if no background exists). For the exclusion line, at that rate, $90\%$ of times, at least one event would be produced, so that if 0 events are observed, the model would be excluded with that level of confidence. 
In Appendix~\ref{sec:analysisdetails} we compare the results of our full log-likelihood ratio test with standard simplified schemes based on the definition of a RoI like the one defined in Fig.~\ref{fig:heatmap} as well as the one more commonly used to derive projections, which just assumes that the ER background can be fully rejected by paying a 50\% efficiency factor on the signal rate~\cite{DARWIN:2016hyl,Aalbers:2022dzr}. 

\begin{figure}[htp!]
    \centering
\includegraphics[width=0.48\textwidth]{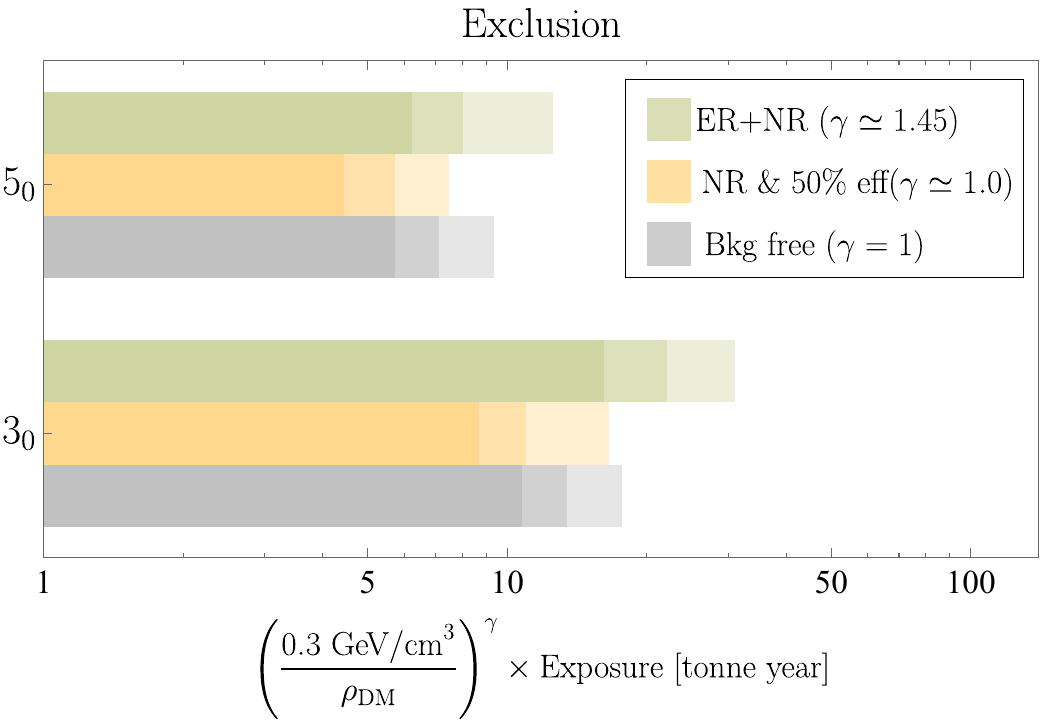} \quad \includegraphics[width=0.48\textwidth]{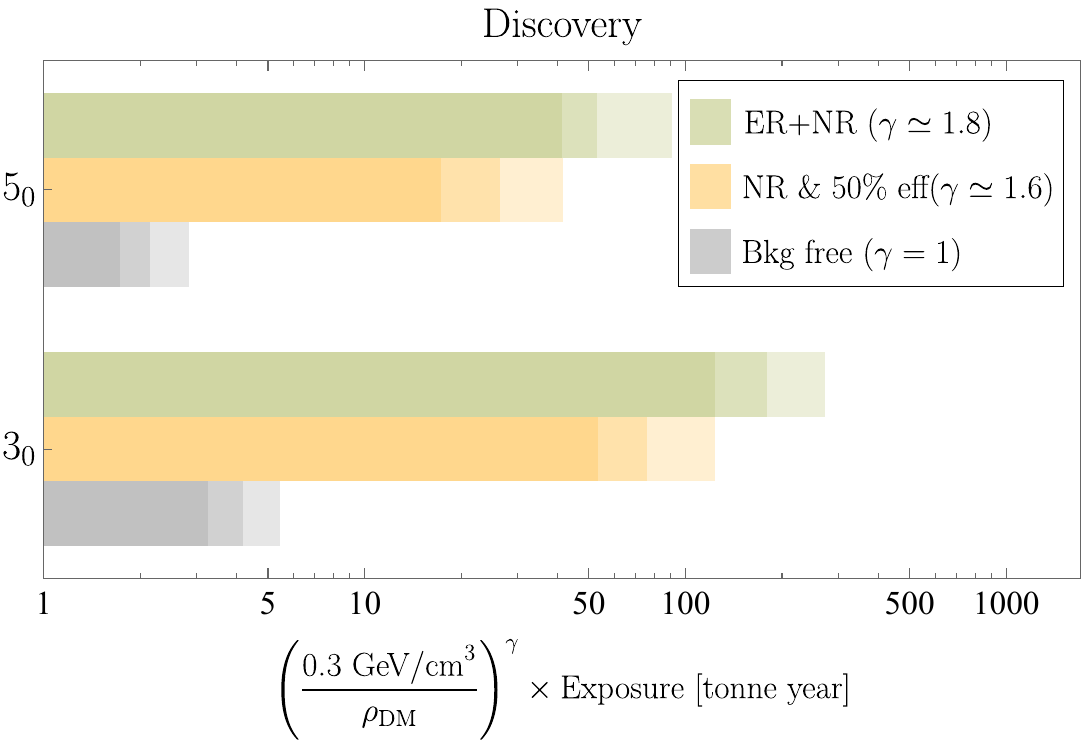}    
\caption{Required exposure for exclusion (upper panel) and discovery (lower panel) for the real 3-plet (pure Wino scenario in Split Supersimmetry~\cite{Arkani-Hamed:2004ymt,Giudice:2004tc,Arkani-Hamed:2004zhs}) and real 5-plet (Minimal Dark Matter candidate~\cite{Cirelli:2005uq,Cirelli:2007xd,Cirelli:2009uv}). In {\color{defgrey} \bf gray} we show the required exposure for exclusion/discovery under the assumption of zero background for comparison. In {\color{yellow35}\bf yellow} we show the required exposure assuming only NR recoils while in {\color{green35}\bf green} the exposure with the addition of ER events.}
    \label{fig:rate_money5and3}
\end{figure}

 We explicitly illustrate this comparison for two well-motivated electroweak (EW) WIMPs in Fig.~\ref{fig:rate_money5and3}, where we employ a simple scaling of the required exposure with the local dark matter (DM) density, in light of the uncertainty in this parameter. In particular, from Fig.~\ref{fig:ExclusionDiscovery}, we fit the required exposure, \(\mathrm{exp}\), as a function of the signal rate \(\mathcal{R}_\chi\) within the range relevant for EW WIMPs using a simple power-law:
\begin{equation}\label{eq:scaling_rho}
\mathrm{exp} \propto \mathcal{R}_\chi^{-\gamma} \propto \rho_{\mathrm{DM}}^{-\gamma},
\end{equation}
where, in the last step, we used Eq.~\eqref{eq:diff_rate}. For example, we find \(\gamma = 1.45\) for the exclusion line and \(\gamma = 1.8\) for the discovery line, while the background-free case scales trivially with \(\gamma = 1\).

Although this work focuses on the implications of the electron recoil ``neutrino fog'' for electroweak WIMPs, the results shown in Fig.~\ref{fig:ExclusionDiscovery} are model-independent. They rely solely on the assumption that the mass of the mediator responsible for the direct detection interaction is larger than the typical momentum transfer, which is of order \(\mathcal{O}(\mu_N v)\), where \(\mu_N\) is the reduced mass of the nucleon--DM system and \(v \sim 10^{-3}\) is the typical DM velocity.

\section{Implications for WIMPs}\label{sec:Signal}
We now map the results of Fig.~\ref{fig:ExclusionDiscovery} in the parameter space of EW WIMPs. These have been classified in Refs.~\cite{Bottaro:2021snn,Bottaro:2022one} under the sole assumption that the total DM energy density today resides in a single DM component which is part of a single EW multiplet. The freeze-out predictions for all the $n$-plets up to the unitarity bound defined in Ref.~\cite{Griest:1989wd} have been derived with the estimated theory uncertainties on the freeze-out masses further refined in Ref.~\cite{Bottaro:2023wjv}. These mass predictions and uncertainties are used throughout this work. We discuss the implication of future direct detection experiments on real WIMPs with zero hypercharge in Sec.~\ref{sec:realWIMP} and complex WIMPs with non-zero hypercharge in Sec.~\ref{sec:complexWIMP}. Further details on the computation of the EW WIMPs signal rate in direct detection are postponed to Sec.~\ref{sec:Signalerror}.

\subsection{Real WIMPs}\label{sec:realWIMP}
For real WIMPs, the DM multiplet is a sigle Majorana fermion whose lagrangian can be  written as
\begin{equation}
\mathscr{L}_{\rm{real}}\supset\frac{1}{2}\chi \left [i\bar{\sigma}^{\mu} \left(\partial_\mu-i g_2 W_\mu^a T^a_\chi\right) -M_{_{\rm{DM}}}\right]\chi\, , \label{eq:WIMP} 
 \end{equation}
with $T^a_\chi$ being the generators in the $n$-th representation of SU(2) which we take to be real so that the hypercharge of the multiplet is automatically zero and $n$ is odd. The first interaction with the Higgs boson appears at dimension seven and it is not important here. 

The required exposure to exclude/discover real WIMPs  at future direct detection detectors is shown in Fig.~\ref{fig:rate_money2} where the associated uncertainty stems from the uncertainty in the WIMP signal rate together with the astrophysical uncertainty on the DM velocity distribution. The uncertainty in the DM local energy density is factored out {\bf using its scaling with the required exposure defined in Eq.~\eqref{eq:scaling_rho}}.
These uncertainties are further discussed in Sec.~\ref{sec:Signalerror}. 

First, we notice that all the real WIMPs with zero hypercharge can be discovered at future large exposure detectors which are planned to have an exposure around one hundred tonne year. Second, as already noticed in Sec.~\ref{sec:setup}, the real WIMP rates are sufficiently larger than the nuclear recoil neutrino rate so that the systematic uncertainties on the latter can be neglected in the estimate of the required exposure. 

Two notable candidates among the real WIMPs are the 3-plet of SU(2) which is well known as the pure Wino scenario in Split Supersymmetry~\cite{Arkani-Hamed:2004ymt,Giudice:2004tc,Arkani-Hamed:2004zhs} and the fermionic 5-plet of SU(2) which is the main candidate of Minimal Dark Matter, the only accidentally stable and fully calculable EW WIMP ~\cite{Cirelli:2005uq,Cirelli:2007xd,Cirelli:2009uv}. In Fig.~\ref{fig:rate_money5and3} we compare the required reach for exclusion/discovery under the assumption of zero background (in gray), the required reach accounting for the ER and NR neutrino fog (in green) and the one that neglects the ER, at the price of a 50\% efficiency~\cite{DARWIN:2016hyl,Aalbers:2022dzr}. As we can see the background-free assumption is grossly under-estimating the required exposure, meaning that the signal region is far from being free of background even for signal rates far above the NR background. The discrepancy becomes larger if one looks at the require exposure for discovery. The standard method of neglecting the ER, and assuming a 50\% efficiency cut lives roughly the right answer, however, it can underestimate or overestimate the required required exposure depending on the signal rate. More details on this comparison are given in Appendix~\ref{sec:analysisdetails}.

\subsection{Complex WIMPs}\label{sec:complexWIMP}
\begin{figure*}[htp!]
    \centering
    \includegraphics[width=0.8\textwidth]{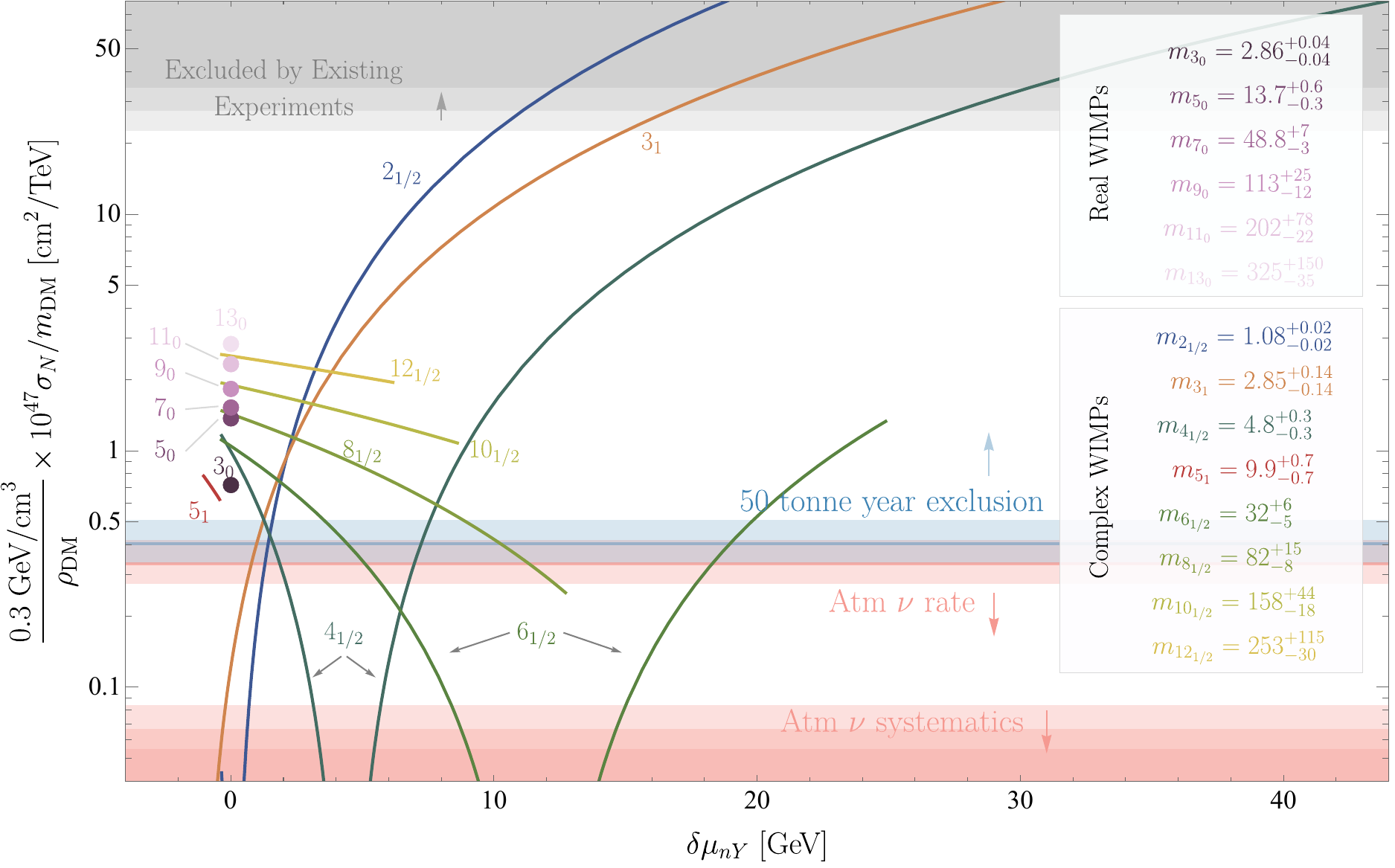}
    \caption{Range of $\sigma_N/m$ spanned by the effective mass splitting defined in \eqref{eq:eff_splitting}, where only the central values are shown. The large dots identify the case where $y_0$ and $y_+$ are the smallest allowed by phenomenology. For real WIMPs, $\sigma_N/m$ is instead fixed. Using that $\sigma_N/\mdm\propto R_\chi/ \eta(v_{\rm min})$ as in \eqref{eq:diffrateWIMP}, the {\color{gray}\bf gray region} corresponds to the excluded rate from existing experiments~\cite{LZ:2022lsv,PandaX-4T:2021bab,XENON:2023cxc}, with the strongest constraint currently from LZ~\cite{LZ:2022lsv}. The {\color{fig34blue}\bf  blue band} shows which range of $\delta\mu_{nY}$ can be excluded with 50 tonne year of exposure. Finally,  the  {\color{cbred}\bf red band} shows when the rate is dominated by neutrino NR due to cancellations in the SI cross-section, while the {\color{cbred}\bf red region} when the rate falls below the uncertainty in the athmospheric neutrino flux. The different shadings and the width of the bands are due to the astrophysical uncertainty encoded in $\eta(v_{\rm min})$. The thermal masses of both real and complex WIMPs derived in Refs.~\cite{Bottaro:2021snn,Bottaro:2022one}, are summarized in the right panel with the uncertainties updated in Ref.~\cite{Bottaro:2023wjv}.}
    \label{fig:mass_splitting}
\end{figure*}
For complex WIMPs, the DM multiplet is a single Dirac fermion with non-zero hypercharge\footnote{Complex WIMPs with $Y=0$ are a simple generalization of the real WIMPs and will not be discussed here. See Ref.~\cite{Bottaro:2022one} for a classification.} whose coupling to the SM gauge bosons can be written as
\begin{equation}\label{eq:lagcomplex}
\mathscr{L}_{\rm{complex}} =\overline{\chi} \left(i\slashed{\partial}-i g_2 \slashed{W}^a T^a_\chi-ig_YY\slashed{B}-m_\chi\right)\chi \ ,
\end{equation}
where slashed quantities are contracted with the Dirac matrices (e.g. $\slashed{\partial}=\gamma^{\mu}\partial_\mu$). At face value, the coupling to the $Z$ boson of the neutral component $\chi_N$ of the EW multiplet would lead to an elastic cross section with nuclei which is grossly excluded by direct detection experiments. 

Thankfully, the Dirac nature of the multiplet and its non-zero hypercharge allow to write a non-renormalizable interactions with the Higgs-boson. For $n$-plets with $Y=1/2$ this is controlled by a dimension 5 operator
\begin{equation}\label{eq:neutral_splitting_v0}
\mathscr{L}_{\rm{split},0}^{Y=1/2}=\frac{y_0}{8\Leff}\left(\overline{\chi}T^a_\chi\chi^c\right)(H^{c\dagger})\frac{\sigma^a}{2} H+\hc\ .
\end{equation}
After EWSB, this operator induces a mixing between $\chi_N$ and $\chi_N^c$. Replacing the Higgs with its VEV $v_h$, $(H^{c\dagger})\frac{\sigma^a}{2} H$ is non-zero only if we pick $\sigma^a=\sigma^+$, so that the new (pseudo Dirac) mass term in the Lagrangian reads
\begin{equation}
\label{eq:neutral_splitting}
\mathscr{L}_m=m_\chi\overline{\chi}_N\chi_N+\frac{\delta m_0}{4}\left[\overline{\chi}_N\chi_N^c +\overline{\chi^c}_N\chi_N\right]\, ,
\end{equation}
with $\delta m_0 = y_0 n v_h^2/(8\Leff)$. The mass eigenstates are Majorana fermions, $\chi_0$ and $\chidm$, with masses $m_0=m_\chi+\delta m_0/2$ and $\mdm=m_\chi-\delta m_0/2$, whose coupling to the $Z$ boson is

\begin{equation}
\label{eq:Z_nonvectorial}
\begin{aligned}
    &\mathscr{L}_Z=\frac{ieY}{\sin\theta_W\cos\theta_W}\overline{\chi}_0\slashed{Z}\chidm\ .
\end{aligned}
\end{equation}
The $Z$-mediated scattering of DM onto nucleons is no longer elastic and the process is kinematically forbidden if the kinetic energy of the DM-nucleus system in the center-of-mass frame is smaller than the mass splitting 

\begin{equation}
    \frac{1}{2}\mu \vrel^2<\delta m_0\ ,\quad \mu=\frac{\mdm m_N}{\mdm+m_N}\ ,
\end{equation}
where $m_N$ is the mass of the nucleus, $\mu$ is the reduced mass and $\vrel$ is DM-nucleus relative velocity. Given the upper bound on the relative velocity $\vrel<v_\text{E}+v_\text{esc}$, the largest testable mass splitting is $\delta m_0^{\max}=1/2\mu (v_\text{E}+v_\text{esc})^2$ which for xenon nuclei gives $\delta m_0^{\max}\simeq 450$ keV. At present, the maximal splitting constrained experimentally is $\delta m_0^{\max,\exp}\simeq 240 \text{ keV}$, given that XENON1T~\cite{XENON:2018voc} analyzed data only for $E_R<40\text{ keV}$. For $M_{\rm DM}\gtrsim 5$ TeV, the dominant lower bound on $\delta m_0$ comes from requiring that the decay $\chi_0\rightarrow \chidm+$SM (where SM=$\gamma, \overline{\nu}\nu$ or $\overline{e}e$) happens well before BBN. Finally, an upper bound on $\delta m_0$ can be derived by requiring that $y_0$ remains perturbative for UV cutoffs $\Lambda_{\text{UV}}\geq 10\mdm$, where our simplifying assumption that the dark sector consists of a single multiplets still holds.

In general, Eq.~\eqref{eq:neutral_splitting} is not sufficient to make complex WIMPs phenomenologically viable. In fact, EW interactions induce at 1-loop mass splittings between the charged and the neutral components of the EW multiplet which in the limit $m_W\ll M_\chi$ are~\cite{Cheng:1998hc,Feng:1999fu,Gherghetta:1999sw}
\begin{equation}
\label{eq:charged_split_gauge}
    \Delta M_Q^{\text{EW}}= \delta_g \left(Q^2+\frac{2YQ}{\cos\theta_W}\right)\ ,
\end{equation}
where  $\delta_g=(167\pm 4)$ MeV and $Q=T_3+Y$ is the electric charge. This implies that negatively charged states with $Q=-Y$ are pushed to be lighter than $\chidm$ by EW interactions with the notable exception of multiplets with maximal hypercharge $|Y_{\max}|=(n-1)/2$ where negatively charged states are not present (e.g. the doublet of $SU(2)$ for $Y=1$). For general complex WIMPs, making $\chidm$ the lightest state requires an extra  dimension 5 operator
\begin{equation}\label{eq:charged_splittinghiggs}
\mathscr{L}_{\rm{split,+}}=- \frac{y_+}{\Leff}\overline{\chi}T^a_\chi\chi\ H^\dagger\frac{\sigma^a}{2} H\ .
\end{equation}

\begin{figure*}[htp!]
    \centering
    \includegraphics[width=0.8\textwidth]{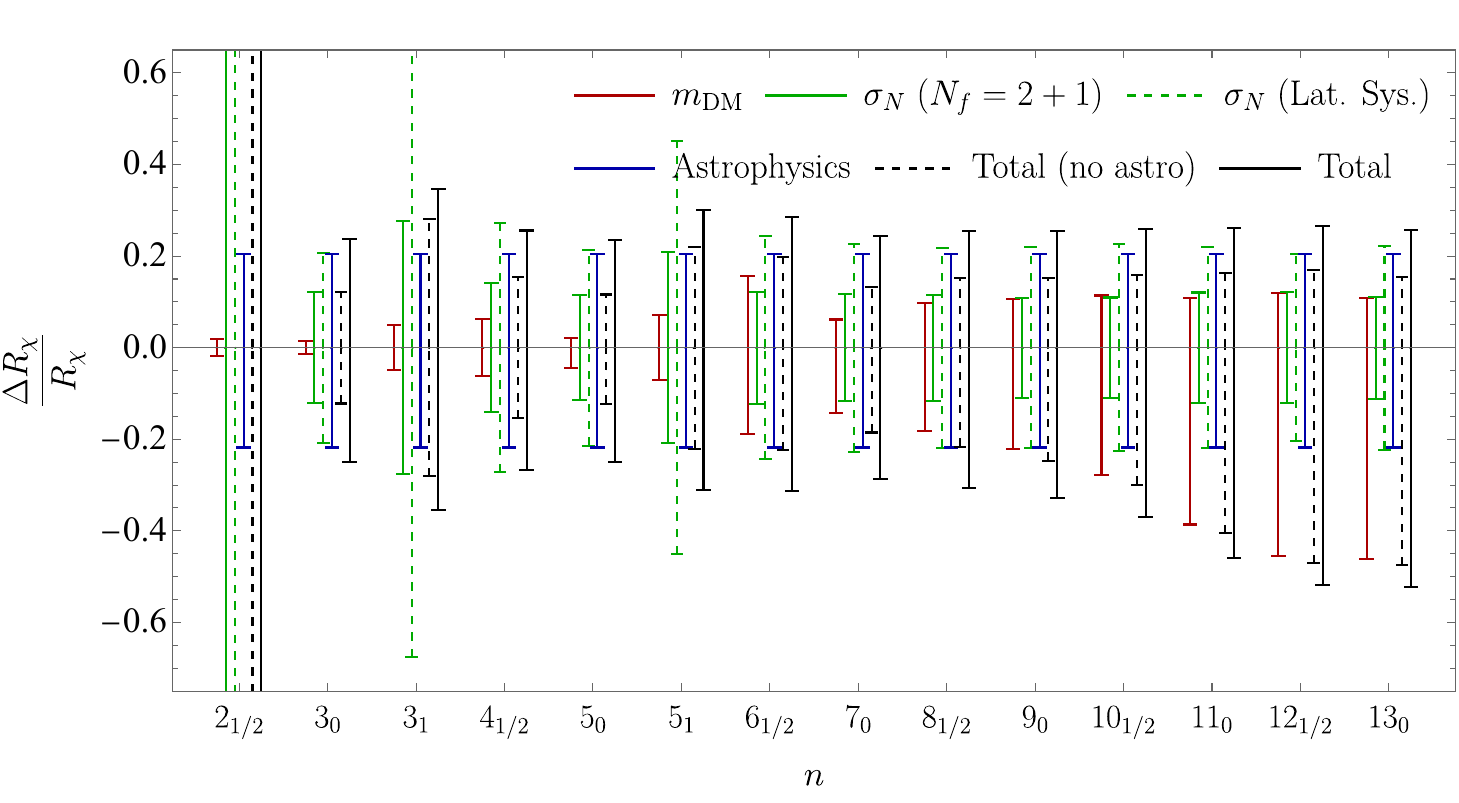}
    \caption{Total relative uncertainty on the EW WIMP signal rate with ({\bf black}) and without ({\bf dashed black}) including the astrophysical uncertainties. The different contributions to the uncertainty come from i) the freeze-out determination of the DM mass ({{\color{fig34red} \bf red}}), ii) the theoretical uncertainty in the spin-independent WIMP cross section ({{\color{cbgreen}\bf green}}), iii) the DM velocity distribution ({{\color{blueWIMP}\bf blue}}). The local energy density is fixed to $\rho_{\rm{DM}}=0.3\,\rm{ GeV/cm}^3$ but all our results can be rescaled for smaller/larger values of $\rho_{\rm{DM}}$. The {{\color{cbgreen}\bf dashed green}} line shows the blow up of the theory uncertainty in the WIMP cross section if we consider the present discrepancy in the lattice data reported in Eqs.\,(\ref{FLAG21}-\ref{FLAG211}) as a theory systematic (see the text for a detailed discussion).}
    \label{fig:deltaR}
\end{figure*}

The viable complex WIMPs parameter space is then determined by two additional free parameters: $y_0$ and $y_+$, whose viable range has been determined in Ref.~\cite{Bottaro:2022one}. A combination of these parameters can be probed in direct detection experiments since the operators in Eq.~\eqref{eq:neutral_splitting} and Eq.~\eqref{eq:charged_splittinghiggs} induce a coupling of the DM to the Higgs boson of the form
\begin{equation}\label{eq:higgs_coupling}
    \mathscr{L}_h=-\frac{\lambda v_h}{2\Lambda_{\rm UV}}\chi_{\rm DM}^2h\ .
\end{equation}
where we define 
\begin{equation}\label{eq:eff_splitting}
\begin{split}
    \frac{\lambda}{\Lambda_{\rm UV}}=\frac{1}{4\Lambda_{\rm UV}}\left(y_+-\frac{n}{2}y_0\right)\equiv -\frac{\delta\mu_{nY}}{v_h^2}\ ,
\end{split}
\end{equation}
and $\delta\mu_{nY}$ is an effective mass splitting which can be written in terms of the physical mass splittings involving neutral and charged states as detailed in Ref.~\cite{Bottaro:2022one}. As we will explicitly show in the next section, the coupling in Eq.~\eqref{eq:higgs_coupling} will contribute to the direct detection cross-section.  

 As shown in Fig. \ref{fig:mass_splitting}, the presence of an independent contribution from the EW one makes it possible to induce large cancellations in the total direct detection rate and bring the cross section for complex WIMPs with $Y=1/2$ well below the neutrino floor. Notable exceptions are the 10-plet and the 12-plet of $SU(2)$ where the EW cross section gets too large with respect to the Higgs-induced one, precluding the possibility of tuning the two contributions against each other. We will elaborate further on this feature in the Sec.~\ref{sec:Signalerror}. 

 A separate mention must be made for the complex WIMPs with $Y=1$. For these WIMPS the equivalent of Eq.~\eqref{eq:neutral_splitting} becomes a dimension 7 operator to ensure  gauge invariance under $U(1)_Y$. Such a higher dimensional operator suppresses the neutral splitting by two extra powers of the UV cut-off with respect to the $Y=1/2$ case. This further suppression leaves only two viable calculable candidates (i.e. with $\Lambda_{\text{UV}}\geq 10\mdm$): the $3_1$ and the $5_1$. For the latter the EW contribution is always larger than the Higgs-induced one making it impossible to tune the total direct detection rate below the neutrino floor. 
 
 Conversely, for the $3_1$ the pure EW contribution happens to be below the neutrino floor because of a peculiar cancellation between the different contributions to the cross section which was first observed for the $2_{1/2}$ (i.e. the pure Higgsino in Supersymmetry~\cite{Hisano:2004pv}). For these two cases the experimental reach will be limited by theoretical systematics on the determination of the atmospheric neutrino fluxes~\cite{Strigari:2009bq} which are not included here. For this reason we leave these two multiplets out of the range of interest of Fig.~\ref{fig:rate_money2}.

\section{Signal rate and uncertainties}\label{sec:Signalerror}

The expression of the differential rate $dR_\chi/dE_R$ of DM collisions per nuclear recoil energy and per unit target mass, whose general form has been shown in Eq.\,(\ref{eq:diff_rate}), can be further simplified in the case of heavy WIMPs. The model-dependence in the differential rate in Eq.\,\eqref{eq:diff_rate} is contained, in fact, in the differential cross-section per unit mass $\mdm^{-1}\mathrm{d}\sigma/\mathrm{d}E_R$. For our WIMP candidates, using the fact that the DM couples in the same way to both proton and neutron (see Appendix \ref{app:theory}), and that $\mdm$ is much larger than the nucleon mass, we get the rather simple expression 
\begin{equation}
\label{eq:diffrateWIMP}
\frac{d R_\chi}{d E_R}= \frac{A^2\rho_{\rm{DM}}}{2\mu_N^2}\frac{\sigma_N}{m_{\rm{DM}}} F^2(E_R)\,\eta(v_{\rm min})\ ,
\end{equation}
where $\sigma_N$ is the spin-independent (SI) cross-section per nucleon, $A\approx131$ is the atomic mass number averaged over xenon isotopes (see Appendix~\ref{app:composition}), $\mu_N$ is the reduced mass of the nucleon-DM system which gets very close to the nucleon mass $m_N=0.939\,\rm{ GeV}$ in the limit $M_{\rm{DM}}\gg m_N$, $F(E_R)$ is the nuclear form factor~\cite{Vietze:2014vsa}. $\eta(v_{\rm \min})$ is the mean inverse speed, defined as
\begin{equation}
 \eta(v_{\rm min}) =  \int_{v>v_{\text{min}}} d^3v \frac{\tilde{f}(\vec{v})}{v}\ .
\end{equation}
The model-dependent part of the rate computation, $\sigma_N/m_{\rm{DM}}$, nicely factorizes from the quantities which instead depends on astrophysics such as the local DM energy density and its velocity distribution.

\begin{figure*}[htp!]
    \centering
    \includegraphics[width=0.75\textwidth]{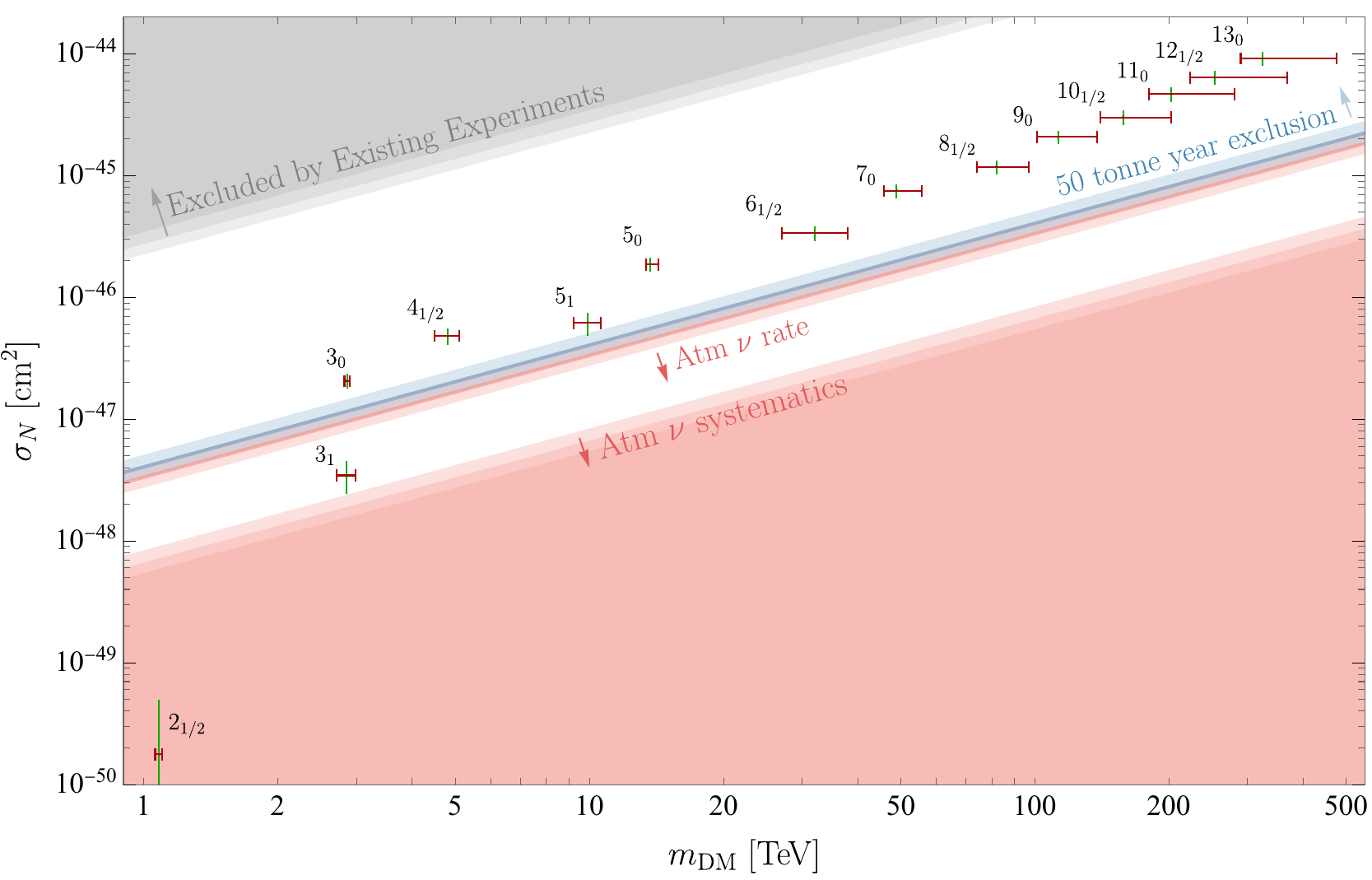}
    \caption{Total SI cross-section per nucleon $\sigma_N$ as a function of the DM mass, $\mdm$. The cross-sections for the complex multiplets are computed choosing $y_0$ and $y_+$ to be the smallest in the allowed range (see Fig.~\ref{sec:complexWIMP}). Fixing  $\rho_{\rm DM}=0.3 ~{\rm GeV/cm^3}$, and recalling that $\sigma_N\propto \mdm R_\chi/ \eta(v_{\rm min})$ from \eqref{eq:diffrateWIMP}, the {\color{gray} \bf gray region } is obtained using the excluded rate from existing experiments~\cite{LZ:2022lsv,PandaX-4T:2021bab,XENON:2023cxc} with the strongest constraint currently from LZ~\cite{LZ:2022lsv}, while the {\color{cbblue}\bf blue band} shows the 50 tonne year exclusion line. The {\color{cbred}\bf red band} shows where the signal is dominated by $\nu$ NR events, while the {\color{cbred}\bf red region} where the rate is smaller than the uncertainty on the athmospheric $\nu$ flux. The different shadings and the width of the bands are due to the astrophysical uncertainty encoded in $\eta(v_{\rm min})$. Finally, the {\color{cbgreen} \bf green} ({\color{cbred} \bf red}) error bands denote the theory uncertainty on WIMP spin-independent cross-section (mass). This figure improves and corrects similar plots in Refs.~\cite{Bottaro:2021snn,Bottaro:2022one}.}
    \label{fig:sigmaplot}
\end{figure*}

The uncertainties in the differential rate can be classified into two categories: i) \emph{theoretical uncertainties} include the uncertainty on the freeze-out prediction of the DM mass summarized in the right panel of Fig.~\ref{fig:mass_splitting} and the uncertainty on cross-sections discussed in Sec.~\ref{sec:theory_err} and showed in Figure \ref{fig:sigmaplot}. These uncertainties can be trated as independent for the DM masses of interest. ii) \emph{Astrophysical uncertainties} include the uncertainty on the local DM density and the one on its velocity distribution and are discussed in Sec.~\ref{sec:astro}. The relative variation on the rate $R_\chi$ induced by each of these sources of uncertainties individually, as well as the total one are summarized in Fig.\,\ref{fig:deltaR} for each $n$-plet.

\subsection{The WIMP cross section and its uncertainty}
\label{sec:theory_err}

In order to assess the theoretical uncertainty on $\sigma_N$, we synthetically discuss the steps relevant for its computation keeping in mind that the SI cross section receives contributions purely from EW loop diagrams for real WIMPs and from a combination of EW diagrams and Higgs-mediated tree-level diagrams for complex WIMPs. In this self-contained summary we revisit the original work of Refs.~\cite{Hisano:2011cs,Hisano:2015rsa,Hill:2014yka,Hill:2014yxa} with further details given in Appendix~\ref{app:theory}.

First of all, we match the UV Lagrangian onto the effective filed theory (EFT) that describes the SI interactions of non-relativistic DM particles with quarks and gluons, obtained by integrating out the EW gauge bosons. The EFT Lagrangian is given by
\begin{equation}\label{LagrUVDD}
\mathscr{L}^{\text{SI}}_{\text{eff}}= f_q \mathcal{O}_{S}^{q} +  f_G \mathcal{O}_{S}^{G} +\!\!\sum_{i=q,G}\left[\frac{g_1^i}{M_\chi} \mathcal{O}_{T_1}^i +\frac{g_2^i}{M_\chi^2} \mathcal{O}_{T_2}^i\right]\, ,
\end{equation}
where the scalar operators are defined as
\begin{equation}
\mathcal{O}_{S}^{q} = \bar{\chi} \chi \mathcal{O}_q\ , \quad
\mathcal{O}_{S}^{G} =\bar{\chi} \chi \mathcal{O}_G\, 
\label{Oscalar}
\end{equation}
with $\mathcal{O}_q\equiv m_q\overline{q}q$ and $\mathcal{O}_G\equiv(\alpha_s/\pi)G_{\mu\nu}^aG^{a\mu\nu}$. The twist-2 operators are
\begin{equation}
\mathcal{O}_{T_1}^i =i\bar{\chi} \partial^{\mu} \gamma^{\nu} \chi\mathcal{O}^i_{\mu\nu}\, ,\quad \mathcal{O}_{T_2}^i =-\bar{\chi} \partial^{\mu} \partial^{\nu}\chi\mathcal{O}^i_{\mu\nu}\, .
\label{Otwist2}
\end{equation}
with $\mathcal{O}^q_{\mu\nu} \equiv \frac{i}{2} \bar{q} \left( D_{\mu} \gamma_{\nu} + D_{\nu} \gamma_{\mu} - g_{\mu\nu}\slashed{D}/2   \right)q$ and $\mathcal{O}^G_{\mu\nu} \equiv G_{\mu}^{a,\rho}G_{\nu\rho}^{a}-(g_{\mu\nu}/4)G_{\rho\sigma}^a G^{a,\rho\sigma}$.

The EW contributions to $f_q$, $f_G$, $g_{1,2}^{q}$ and $g_{1,2}^{G}$ at NLO in $\alpha_s$ are discussed in depth in App.\,\ref{WClimit}, where their full expressions in the limit $M_{\rm{DM}}\gg m_W$ are given as obtained by Ref.~\cite{Hisano:2015rsa} at NLO in $\alpha_s$. Neglecting the NLO corrections, the results of \cite{Hisano:2015rsa} are equivalent to those of \cite{Hill:2014yxa}. The contributions from the Higgs exchanges affect $f_q$ and $f_G$ and can be obtained by integrating out the Higgs in the operator in Eq.~\eqref{eq:higgs_coupling}. The  general expressions for $f_q$ and $f_G$ for complex WIMPs are then\footnote{In the original version of this manuscript, the Higgs contribution to $f_G$ was incorrectly given as $\delta\mu_{nY}/(16 m_h^2v_h^2)$. The authors are grateful to Stéphane Lavignac for bringing this error to our attention.}
\begin{equation}\label{eq:Wilson_coeff_higgs}
    f_q=f_q^{\rm EW}-\frac{\delta\mu_{nY}}{2m_h^2v^2_h},\quad f_G=f_G^{\rm EW}+\frac{\delta\mu_{nY}}{24m_h^2v^2_h}\ ,
\end{equation}
where $f_q^{\rm EW}$ and $f_G^{\rm EW}$ are the pure EW contributions, $v_h$ indicates the Higgs VEV and $\delta\mu_{nY}$ is defined in Eq.~\eqref{eq:eff_splitting}.

From Eq.\,(\ref{LagrUVDD}) one can derive explicitly the expression of the SI elastic cross-section per nucleon, which takes the rather simple form in the limit $m_\chi\gg m_N$:
\begin{equation}
\label{elasticSIDD}
\sigma_{\text{SI}}^{\text{EW}} \simeq \frac{4}{\pi} m_N^4 \vert k_S^{\text{EW}}+k_T^{\text{EW}} \vert^2,
\end{equation}
where $m_N$ is the nucleon mass, while $k_S^{\text{EW}}$ and $k_T^{\text{EW}}$ are the scalar and twist-2 amplitudes, respectively, whose expressions are defined below. 
 
 For the twist-2 amplitude, the hadronic matrix elements of the operators defined below Eq.\,(\ref{Otwist2}) can be written as 
\begin{align}
&\langle N(p)\vert \mathcal{O}^q_{\mu\nu} \vert N(p)\rangle= \Pi_{\mu\nu}  (q(2,\mu) + \bar{q}(2,\mu))\\
&\langle N(p)\vert \mathcal{O}^G_{\mu\nu} \vert N(p)\rangle=-\Pi_{\mu\nu} g(2,\mu)\ ,
\end{align}
where $\Pi_{\mu\nu}=\frac{p_{\mu}p_{\nu}}{m_N} - \frac{ m_N}{4} g_{\mu\nu}$ is the transverse traceless projector and $q(2,\mu)$, $\bar{q}(2,\mu)$, and $g(2,\mu)$ are the second moments of the PDFs of quark, antiquark and gluon in the nucleon defined at the generic scale $\mu$. The values of the PDFs can be directly evaluated at the EW scale $m_Z$ and, thus, the twist-2 amplitude can be simply written as
\begin{align}\label{eq:kTEW}
k_T^{\text{EW}} \!&= \! \frac{3}{4}\! \sum_q (q(2,m_Z) + \bar{q}(2,m_Z))(g_1^{q}(m_Z)+g_2^{q}(m_Z))\notag\\
&-\frac{3}{4} g(2,m_Z) (g_1^{G}(m_Z)+g_2^{G}(m_Z))\, ,
\end{align}
where the coefficients $g_{1,2}^{q}$ and $g_{1,2}^{G}$ have been introduced in Eq.\,(\ref{LagrUVDD}) and the sum runs over all the active quarks ($q = u,\,d,\,s,\,c,\,b$). In this work we have used the same numerical values of the second moments of the PDFs adopted in Ref.~\cite{Hisano:2015rsa} originally extracted by the CTEQ-Jefferson Lab collaboration in Ref.\,\cite{Owens:2012bv} and reported in Table~\ref{tab:PDFmoments} together with their uncertainty. 

For the scalar amplitude, the hadronic matrix elements of the scalar operators for quarks and gluons defined below Eq.\,(\ref{Oscalar}) are defined as
\begin{align}\label{eq:hadmat}
&\langle N|  \mathcal{O}_q |N\rangle = m_N f_{Tq}\, ,\quad \langle N\vert\mathcal{O}_G \vert N\rangle\equiv m_N\,f_{TG}\, ,
\end{align}
where $f_{TG}=\frac{4\,\alpha_s^2}{\pi\,\beta(\alpha_s)}\left[ 1- (1-\gamma_m) \sum_q f_{Tq}\right]$ as a consequence of the sum rule derived from the trace anomaly of the QCD energy momentum tensor  \cite{Shifman:1978zn} to which both the gauge coupling beta-function $\beta(\alpha_s) \equiv \mu\,d\alpha_s/d\mu$ and the anomalous dimension of the quarks $\gamma_m\,m_q \equiv \mu\,dm_q/d\mu$ contribute. The value of $f_{Tq}$ are extracted from lattice simulations at the hadronic scale, $\mu_{\rm had}\approx 1$ GeV, where three quarks are active. As a consequence the scalar amplitude is
\begin{equation}
\label{kSEW}
k_S^{\text{EW}} =  \sum_{q} f_q(\mu_{\rm had}) f_{Tq}+ f_G(\mu_{\rm had})f_{TG}\ , 
\end{equation}
where $q = u,\,d,\,s$, and $f_q(\mu_{\rm had})$ and $f_G(\mu_{\rm had})$ are the scalar Wilson coefficients in Eq.~\eqref{LagrUVDD} run from the EW to the hadronic scale. The details of the running down to the hadronic scale are summarized in Appendix \ref{app:running} which follows the work of Ref.~\cite{Hill:2014yxa}. Notice that, the scalar and twist-2 operators do not mix under the RG flow at any order in perturbation theory so it is consistent to evaluate these contributions at different scales.

The hadronic matrix elements in Eq.~\eqref{eq:hadmat} can be extracted from quantities which are computed on the lattice fixing $m_N=0.939\,\rm{ GeV}$. In particular we have 
\begin{equation}
\frac{\sigma_{\pi N}}{m_N}\equiv (f_{Tu}+f_{Td}),\qquad \frac{\sigma_{s}}{m_N}\equiv f_{Ts} \,,    
\end{equation}
where, for $\sigma_{\pi N}$ and $\sigma_s$ we have used the averages of the lattice results performed by the FLAG Collaboration~\cite{FLAG:2021npn}, which considers separately the cases of $N_f=2+1+1$ and $N_f=2+1$ dynamical quarks in the lattice simulations. The FLAG averages for $N_f=2+1$ are obtained by combining three different lattice computations\,\cite{BMW:2011sbi, Durr:2015dna, Yang:2015uis} for $\sigma_{\pi N}$ and five results from Refs.\,\cite{BMW:2011sbi, Durr:2015dna, Yang:2015uis, Freeman:2012ry, Junnarkar:2013ac} for $\sigma_s$, all of them passing the FLAG rating criteria, and read
\begin{equation}
\begin{split}
\label{FLAG21}
&\sigma_{\pi N}=39.7\pm 3.6\,{\rm MeV}\,, \\
&\sigma_{s}=52.9\pm 7.0\,{\rm MeV}\,.
\end{split}
\end{equation}
The FLAG results for $N_f=2+1+1$ are taken from the single study of Ref.~\cite{Alexandrou:2014sha} for $\sigma_{\pi N}$ and from the one of Ref.~\cite{Freeman:2012ry} for $\sigma_s$, both passing the FLAG rating criteria, and are
\begin{equation}
\begin{split}
\label{FLAG211}
&\sigma_{\pi N}=64.9\pm 13.3\,{\rm MeV}\,, \\
&\sigma_{s}=41.0\pm 8.8\,{\rm MeV}\,,
\end{split}
\end{equation}
The values of $\sigma_s$ computed in the two flavor schemes agrees with each other while an important tension exists among the $N_f=2+1+1$ and $N_f=2+1$ FLAG averages of $\sigma_{\pi N}$. We focus on the FLAG lattice results in the $N_f=2+1$ flavor scheme for all the results in the main text of this paper. The necessary ingredients to recast the results for $N_f=2+1+1$ are given in Appendix~\ref{app:theory}. 

All in all, the error on cross section per nucleon $\sigma_N$ is affected by three sources of uncertainty: $i)$ the running of the scalar quark and gluon coefficients (including threshold corrections from heavy quarks) which is truncated at a given order in perturbation theory $ii)$ The lattice values of the matrix elements and $iii)$ the uncertainties on the nucleon PDFs. 

At fixed lattice flavor scheme these uncertainties contribute equally to the current theoretical uncertainty on the signal cross section which is roughly 10\% as shown in Fig.~\ref{fig:deltaR}. However, if we account for the discrepancy between the central values of $\sigma_{\pi N}$ in the 2 flavor schemes as a systematical uncertainty on $\sigma_{N}$ then its theoretical error blows up as shown by the green dashed line in Fig.~\ref{fig:deltaR}. For this reason it is of primary importance to resolve this discrepancy. In Appendix~\ref{app:latticeerror} we further comment on the possible origin of this tension which resides in the challenge of computing the hadronic matrix element on the lattice. 

In order to understand the results in Fig.~\ref{fig:deltaR} it is crucial to remember that the relative theory error on the signal rate gets larger in fine-tuned configurations where the contributions from the different operators to the total scattering cross section partially (or totally) cancel among each others resulting in a smaller rate then the one naively expected.

For the $2_{1/2}$ and the $3_1$ multiplets the cancellation occurs for the purely EW cross section (i.e. at $\delta\mu_{nY} \simeq 0$ as defined in Eq.~\eqref{eq:eff_splitting}). In these cases the twist-2 contributions and the scalar ones are the ones cancelling among each other. The cancellation is such that the total cross section drops below the neutrino floor even thought the contributions from the single operators would be above it. Clearly these large cancellations are very sensitive to the theoretical uncertainties as shown in Fig.~\ref{fig:deltaR}. For example for the $2_{1/2}$ ($3_1$) the cross section induced by the scalar contribution only is $(3.4\pm 0.1)\times 10^{-47}\rm{ cm^2}$ $((3.1\pm 0.1)\times 10^{-46}\,\rm{cm}^2)$, the one from the twist-2 contribution only is $(3.6\pm 0.1)\times 10^{-47}\,\rm{ cm^2}$ ($(2.4\pm 0.1)\times10^{-46}\, \rm{cm}^2$) and the resulting total cross section is $1.8^{+2.8}_{-1.8}\times 10^{-50}\,\rm{ cm^2}$ ($(3.5\pm1.0)\times 10^{-48}\,\rm{cm}^2$). The cancellation for the $2_{1/2}$ can be perfect and as a consequence the relative theoretical uncertainty blows up completely, while for the $3_1$ the cancellation is only partial and the relative uncertainty is around 30\%.\footnote{The result here disagrees with the one presented in Ref.~\cite{Bottaro:2022one}, where the $3_1$ scattering cross section was above the neutrino floor while the $5_1$ one was below. This discrepancy can be traced back to the more careful treatment of the lattice inputs and their running implemented in this work.} 

\begin{figure}[htp!]
    \centering
    \includegraphics[width=0.475\textwidth]{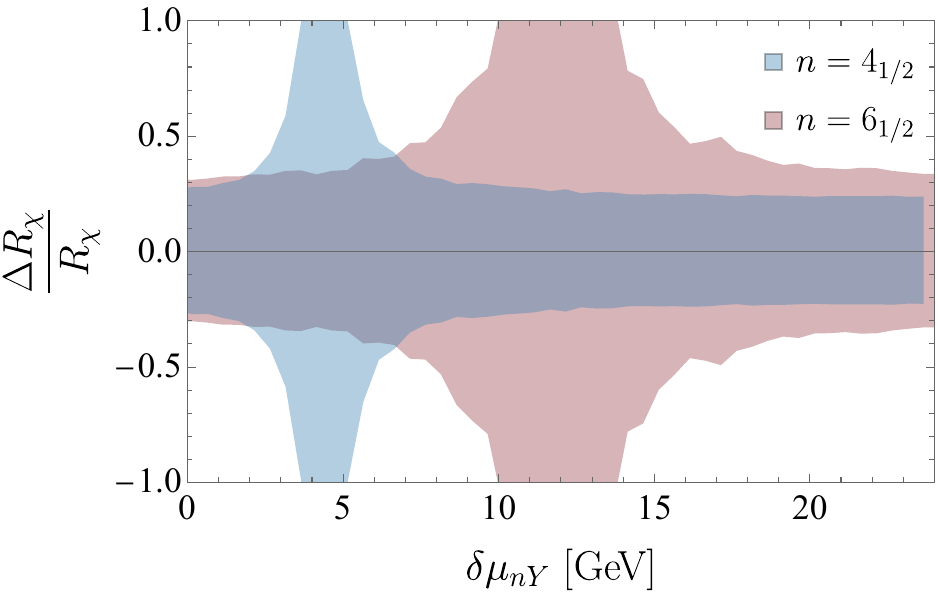}
    \caption{Total relative uncertainty on the EW WIMP signal rate as a function of the effective mass splitting $\delta \mu_{nY}$ defined in Eq.~\eqref{eq:eff_splitting}. The $n=4_{1/2}$ is shown in {\color{cbblue}\bf blue} and the $n=6_{1/2}$ in {\color{fig34red}\bf red}. As expected, the error blows up when the Higgs and the EW contributions cancel out.}
    \label{fig:tuning46}
\end{figure}

For the $4_{1/2}$ and the $6_{1/2}$ multiplets the cancellation happens between the EW contributions and the Higgs one for some non-zero value of $\delta\mu_{nY}$ as shown in Fig.~\ref{fig:tuning46}. Interestingly, this tuning becomes impossibile for larger complex representations because the EW contribution to the cross section grows faster with $n$ than the Higgs one.

\subsection{Astrophysical uncertainties}\label{sec:astro}

The first source of astrophysical uncertainty comes from the knowledge of the local DM energy density, $\rho_{\rm DM}$. This quantity can be experimentally determined in a variety of ways as reviewed in Ref.~\cite{Read:2014qva}. The PDG\,\cite{PDG} quotes the range (0.2 - 0.6) GeV/cm$^3$ for recent results from global fits of different measurements and (0.3 - 1.5) GeV/cm$^3$ for recent analyses of stellar data from the \emph{Gaia} satellite. Since in our study both the rate and the predicted exposures have a simple scaling with $\rho_{\rm DM}$, we have used as a reference value $\rho_{\rm DM}=0.3 {\rm ~GeV/cm^3}$ and made the scaling explicit in all the plots.

\begin{figure}[htp!]
    \centering
    \includegraphics[width=\linewidth]{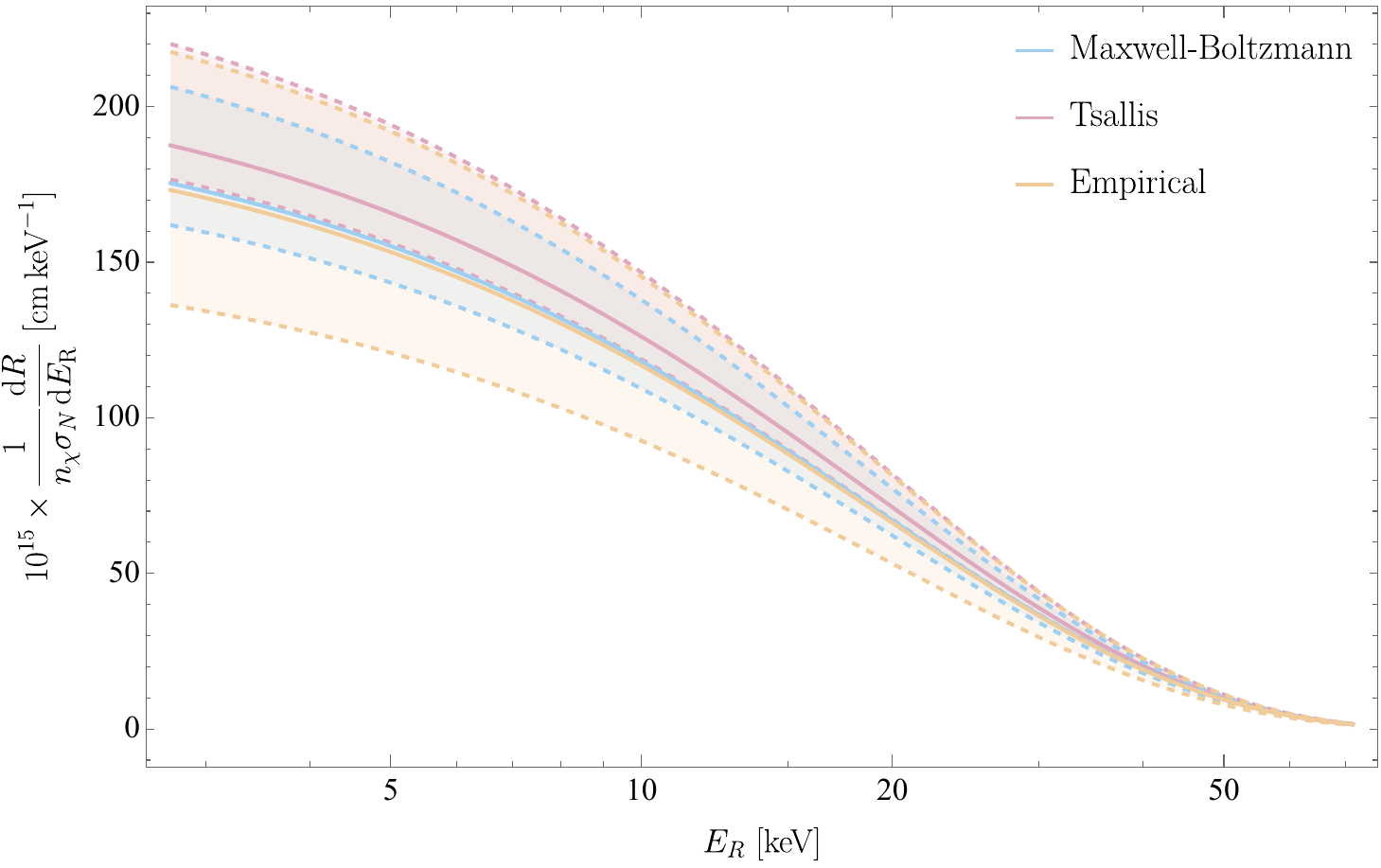}
    \caption{Differential rate, normalized to the DM number density times the SI cross-section per nucleon $\sigma_N$, as a function of the recoil energy for three different velocity distributions: the {\color{Boltzmann} \bf Maxwell-Boltzmann} distribution in Eq.~\eqref{eq:MB}, the {\color{Tsallis} \bf Tsallis} distribution in Eq.~\eqref{eq:TS} and the {\color{Empirical}\bf empirical} distribution in Eq.~\eqref{eq:E}. The uncertainty is extracted by scanning over the allowed range of astrophysical parameters as detailed in Table~\ref{tab:astro_par}.}
    \label{fig:rate_distro}
\end{figure}

The second source of astrophysical uncertainty is related to the choice of the functional structure of the galactic-frame velocity distribution of DM $f(\vec{v})$, which is related to corresponding distribution $\tilde{f}(\vec{v})$ into the Earth's frame as $\tilde{f}(\vec{v}) = f(\vec{v}+\vec{v}_\oplus)$, $\vec{v}_\oplus$ being the Earth velocity with respect to the galactic center. 
A common choice in literature is represented by a Maxwell-Boltzmann velocity distribution \cite{Drukier:1986tm, Lisanti:2016jxe}. A more realistic form of the galactic-frame velocity distribution of DM can be directly extracted from $N$-body simulations. This flourishing field of research has highlighted many interesting deviations from the standard Maxwell-Boltzmann velocity distribution.

To assess the impact of the uncertainty in the velocity distribution in our study we have used three different galactic-frame velocity distributions of DM:
\begin{itemize}
    \item the truncated Maxwell-Boltzmann distribution 
    \begin{equation}\label{eq:MB}
    f_{\rm MB}(\vec{v})\propto  e^{-|\vec{v}|^2/v_0^2}\theta(v_{\rm esc}-|\vec{v}|)\ ,    
    \end{equation}
    \item the Tsallis distribution \cite{Hansen:2005yj, Vergados:2007nc, Ling2010}
     \begin{equation}\label{eq:TS}
    f_{\rm Ts}(\vec{v})\propto \left[1-(1-q)\frac{|\vec{v}|^2}{v_0^2}\right]^{1/(1-q)}\theta(v_{\rm esc}-|\vec{v}|)\ ,
     \end{equation}
    \item the empirical distribution derived in Refs.~\cite{Mao:2012hf, Mao:2013nda}
   \begin{equation}\label{eq:E}
    f_{\rm Emp}(\vec{v})\propto \ e^{-|\vec{v}|/v_0}(v_{\rm esc}^2-|\vec{v}|^2)^p \theta(v_{\rm esc}-|\vec{v}|)\ ,
    \end{equation}
\end{itemize}
where $v_{\rm esc}$ and $v_0$ are the escape velocity and the circular velocity of the local system of rest (LSR)\footnote{The LSR is the rest frame at the position of the Sun that would move with a constant circular velocity if the Milky Way were azimuthally symmetric.}, respectively. 

The astrophysical uncertainties associated to the DM velocity distribution have then been computed by scanning over the parameters of the three galactic-frame velocity distributions above. In the Tsallis model, the parameter $q$ is determined from $v_{\rm esc}$ and $v_0$ as $q=1-v_0^2/v_{\rm esc}^2$, while in the empirical model, following \cite{Mao:2012hf, Mao:2013nda, Radick:2020qip}, we vary $p\in [0,3]$.  In Table~\ref{tab:astro_par}, we summarize the values chosen for the various parameters of the velocity distributions. The central values match the recommended values in \cite{Baxter:2021pqo}, while the error bands are conservatively chosen as explained above.

\begingroup
\setlength{\tabcolsep}{10pt}
\renewcommand{\arraystretch}{1.5}
\begin{table}[h!]
    \centering
     \vspace{0.4cm}
    \begin{tabular}{c| c}
        $v_{\rm esc}$ & $544^{+56}_{-94}$ km/s \cite{Smith:2006ym}\\
        \hline
         $v_0$ & $238^{+42}_{-38}$ km/s \cite{Kerr:1986cei}\\
         \hline
         $v_\oplus$ & $254^{+3}_{-37}$ km/s \cite{Lee:2013xxa,Baxter:2021pqo}\\
    \end{tabular}
    \caption{Range of the relevant astrophysical parameters.}
    \label{tab:astro_par}
\end{table}
In Fig.\,\ref{fig:rate_distro}, we show the differential rates for the different velocity distributions after having marginalized over all the astrophysical parameters. The quoted astrophysical uncertainty has been obtained by taking the combination of velocity distribution and astrophysical parameters that maximize and minimize the rate, using as a reference rate the one given by the Maxwell-Boltzmann distribution with the central values of Table \ref{tab:astro_par}.\\

\endgroup

\section{Conclusions}\label{sec:conclusions}

The question ``is the DM electroweak?'' prompted a vast experimental program that ranges from large telescopes hunting for DM annihilation signals~\cite{Fermi-LAT:2016uux,HESS:2022ygk,VERITAS:2017tif,MAGIC:2021mog,HAWC:2019jvm,CTAO:2024wvb}, to high-energy colliders producing DM directly~\cite{Accettura:2023ked,Aime:2022flm,FCC:2018byv,CEPCStudyGroup:2018rmc} and large scale underground detectors hunting for the DM elastic collisions with the detector nuclei~\cite{Mount:2017qzi,XENON:2020kmp,DARWIN:2016hyl,Aalbers:2022dzr,PandaX:2024oxq,LZ:2022lsv,DARWIN:2016hyl,Baudis:2024jnk}. While most of the theoretical work so far focused on few particularly motivated EW candidates~\cite{Baumgart:2023pwn,Rodd:2024qsi,Rinchiuso:2020skh,Han:2020uak,Capdevilla:2021fmj,Cirelli:2015bda,Krall:2017xij,Bottaro:2021srh,Rinchiuso:2018ajn,Fan:2013faa} it seems plausible that such a sharp question about the nature of DM can be answered in general, for any allowed EW representation.  

In this spirit, we updated the theoretical predictions and uncertainties of the expected signal rate in large scale xenon experiments for all the EW WIMPs and derived the expected sensitivity of future xenon experiments.

All in all, Fig.~\ref{fig:rate_money2} and Fig.~\ref{fig:sigmaplot} show that an exposure of 50 (300) tonne year is enough to exclude (discover) all the real EW WIMPs and severely constrain the parameter space of the complex EW WIMPs with non-zero hypercharge. In Fig.~\ref{fig:rate_money5and3} we updated the required exposure to exclude (discover) the EW triplet (the pure Wino scenario in Supersymmetry) and the EW fiveplet (the Minimal Dark Matter condidate). For these candidates the direct detection projections will complement the prominent role of future indirect detection searches at CTA~\cite{Rinchiuso:2020skh,Baumgart:2023pwn}. Conversely, direct detection seems the best way to probe heavy EW WIMPs, where indirect dection is expected to loose sensitivity~\cite{Bottarotoappear}.  

Fig.~\ref{fig:mass_splitting} shows that the pure EW cross section of the two lightest complex WIMPs (the doublet with hypercharge 1/2 and the triplet with hypercharge 1) lies well below the nuclear recoil neutrino fog while for heavier multiplets possible cancellations between EW and Higgs contributions to the SI cross-section can make the signal rate drop below the nuclear recoil neutrino fog. Interestingly, these cancellations cannot take place for the heavy multiplets and for the fiveplet with hypercharge 1 which can then be excluded (discovered) with a 40 (400) tonne year exposure. 

In our assessment of the expected reach we assumed that all the present noise sources in xenon detectors will be mitigated but we included both nuclear recoil neutrino events and electron recoil neutrino events. The latter background leaks in the signal region because of the imperfect discrimination between electron recoils and nuclear recoils. The current discrimination power makes this background important for EW WIMPs and we showed that the full two-dimensional information of the dual-phase Time Projection Chambers is crucial to optimally reject it, while keeping most of the WIMP signal. For the theoretically motivated cases of the triplet and the quintuplet with zero hypercharge we showed in Fig.~\ref{fig:rate_money5and3} how the inclusion of this background and our likelihood analysis compare to previous simplified treatments~\cite{DARWIN:2016hyl,Aalbers:2022dzr}. 

Along the way, we highlighted the need of further studies to better determine the experimental backgrounds and to reduce the theoretical uncertainties on the WIMP cross section. Regarding the experimental backgrounds, despite much attention given to the systematic uncertainty in the expected number of interactions between neutrinos and the xenon nuclei, \textit{the NR neutrino fog}, we noticed that the uncertainties in detector response to neutrinos interacting with electrons, \textit{the ER neutrino fog}, have garnered far less attention, and are crucial in distinguishing signal from ER background. A robust estimate of those uncertainties requires dedicated experimental works.

As far as the theory predictions are concerned, we identified a significant disagreement between nuclear matrix elements computed in the lattice using $N_f=2+1$ and $N_f=2+1+1$ flavour schemes. If this disagreement is taken as a systematic uncertainty on the WIMP signal rate, the lattice uncertainty becomes of the same order if not larger than the astrophysical uncertainty stemming from the unknown DM velocity distribution in the Galaxy as shown in Fig.~\ref{fig:deltaR}. In Appendix~\ref{app:theory} we identified the possible sources of this discrepancy whose solution requires further  dedicated lattice studies.

\section{Acknowledgements}
We thank Laura Baudis for interesting discussions that prompted this study. We also thank Ranny Budnik, Micha Weiss, and Scott Haselschwardt for providing useful comments and suggestions regarding detector physics. We thank Silvano Simula for useful discussions on the non-perturbative determination of the sigma terms on the lattice. We thank Michele Redi for suggesting Fig.~\ref{fig:rate_money5and3}. We further thank Laura Baudis, Scott Haselschwardt and Ranny Budnik for comments on the draft, and Stéphane Lavignac for pointing out an error in the original version of the manuscript. We thank the Galileo Galilei Institute for hospitality during the development of this study. This work was performed in part at Aspen Center for Physics, which is supported by National Science Foundation grant PHY-2210452. SB is supported by the Israel Academy of Sciences and Humanities \& Council for Higher Education Excellence Fellowship Program for International Postdoctoral
Researcher.

\bibliographystyle{JHEP}
\bibliography{wimp.bib}

\clearpage
\newpage

\appendix

\section{Likelihood Analysis and Cut and Count Comparisons}\label{sec:analysisdetails}

\begin{figure}[htp!]
    \centering
    \includegraphics[width=\linewidth]{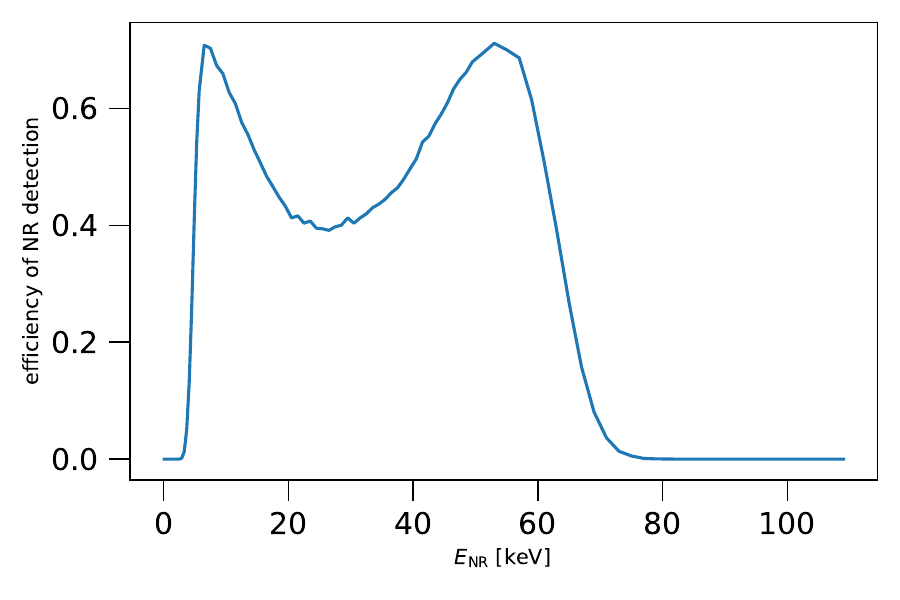}
    \caption{Efficiency of our optimal mask in retaining NR events as a function of NR event energy. A simulation of nuclear-recoil events of a well defined energy was done, and the y axis represents the fraction of events that remained within our optimal cut (defined in the text).}
    \label{fig:NREff}
\end{figure}

\begin{figure}[htp!]
    \centering
    \includegraphics[width=\linewidth]{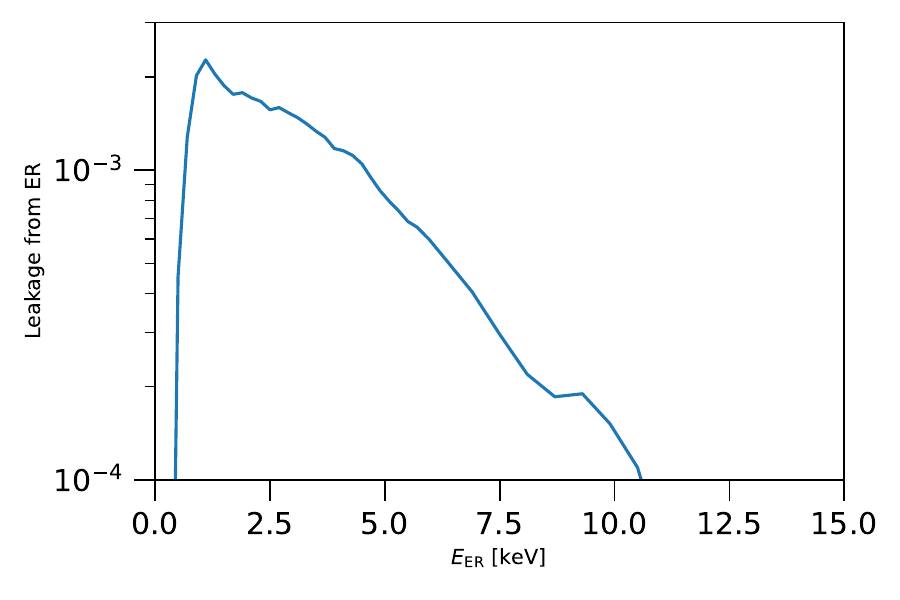}
    \caption{Leakage for ER from our optimal mask. A simulation of electron-recoil events of a well defined energy was done, and the y axis represents the fraction of events that penetrated the optimal cut (defined in the main text).}
    \label{fig:EREff}
\end{figure}

Here we first describe how the optimal mask showed in Fig.~\ref{fig:heatmap} was derived in the cut-and-count scheme. To define the optimal mask, we first removed bins where no signal was found in the $10^7$ simulations ran. We then used the asymptotic estimate of the discovery projection test (see e.g.~Ref.~\cite{Bhattiprolu:2020mwi}), to find which bin-choice is optimal to receive the best projected reach. While the approximate form is in principle only valid for large-backgrounds, using it allows to find a single mask that does not depend on neither the total exposure nor signal normalization. The efficiency in retaining NR events can be seen in Fig.~\ref{fig:NREff}. Conversely the leakage from ER events as a function of energy can be seen in Fig.~\ref{fig:EREff}. Both plots were slightly smoothed for aesthetic purposes (neither is used for any conclusion, and both are used to illustrate the main features of the mask), see \href{https://github.com/ItayBM/NeutrinoFogs}{github} for the exact procedure used.


\begin{figure*}[htp!]
    \centering
    \includegraphics[width=0.48\linewidth]{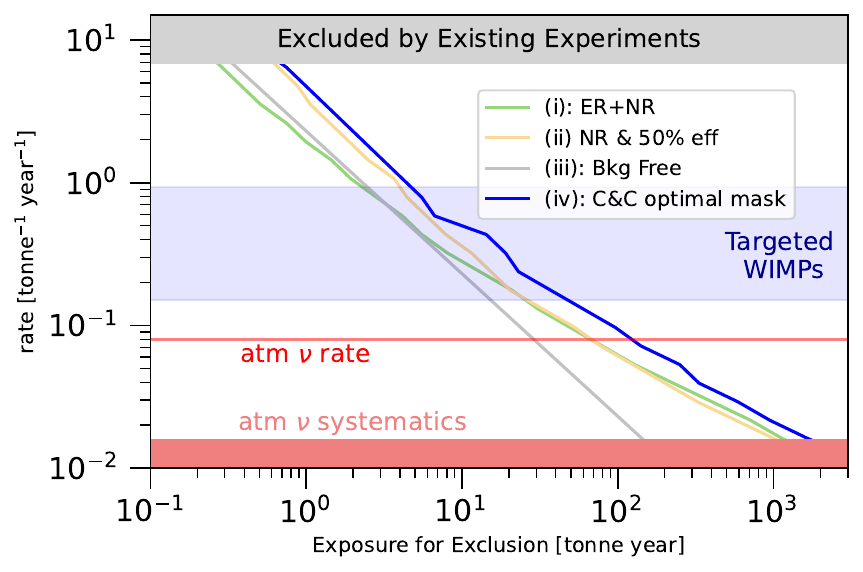}\hfill
    \includegraphics[width=0.495\linewidth]{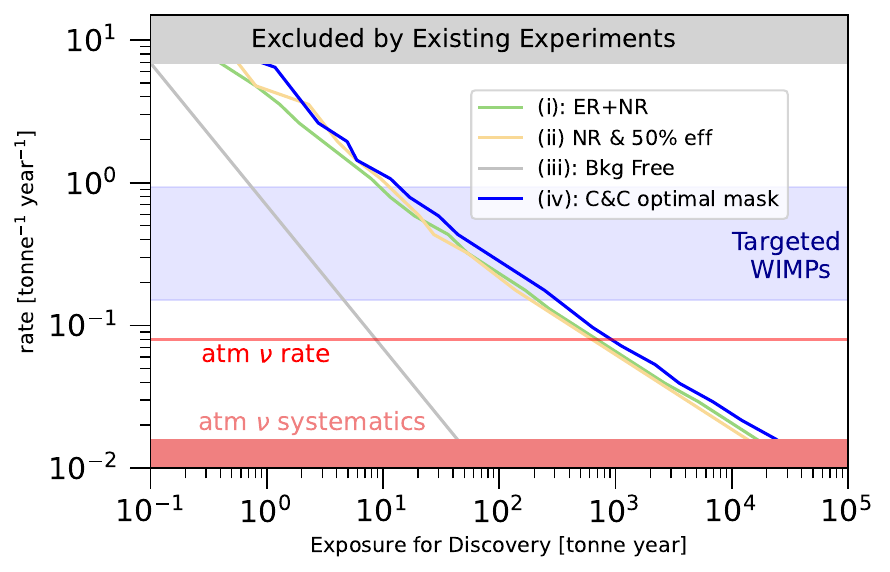}
\caption{Comparing the projected reach for 90\% CL exclusion (top) and a $5\sigma$ discovery (bottom) of different approaches. {\color{green35}\bf Case (i)} is utilizing the full LLR descibed in Sec.~\ref{sec:results}, with the background accounting for both NR and ER neutrino scattering events. {\color{yellow35}\bf Case (ii)} assumes that only NR neutrino events contribute to the background rate, assigning a 50\% efficiency cut on the signal rate (simply multiplying the required exposure by 2). {\color{defgrey} \bf Case (iii)} is the simple background-free estimate. {\color{blueWIMP} \bf Case (iv)} is a cut-and-count analysis after applying our optimal mask. See text for further discussion.}
    \label{fig:expofratecomparison}
\end{figure*}

In Fig.~\ref{fig:expofratecomparison} we compared different analysis methods and different assumptions on the background compositions: 
\begin{enumerate}
    \item[i)] The full 2D LLR test defined in Sec.~\ref{sec:results} that utilizes the full information provided by the dual-phase Time Projection Chambers to reject the background of both neutrino nuclear recoils and electron recoils (see Sec.~\ref{sec:neutrinobkd}).
    \item[ii)] A 2D LLR analysis that only has NR background events. We assume a 50\% efficiency cut, to conform with the standard method of estimating sensitivity near the neutrino floor~\cite{DARWIN:2016hyl,Aalbers:2022dzr}.
    \item[iii)] A simple background-free estimate of the sensitivity.
    \item[iv)] The cut and count analysis, for the ROI selected by our mask in Fig.~\ref{fig:heatmap}. We use the same test-statistics we had for the full LLR, albeit with a single bin for our projected sensitivities.
\end{enumerate}

We compare here below the different methods discussing separately exclusion and discovery. First, for exclusion, we note that at large rates (as seen already in Fig.~\ref{fig:ExclusionDiscovery}), the full LLR reproduces the background free result. As the signal rate becomes lower, the full LLR starts becoming more similar to the standard (ii) method, which is reasonable, since it is known to produce the right results around the neutrino floor. The cut-and-count analysis is always outperformed by the full LLR, as expected, since it neglects a fair amount information. At low rates however, the cut-and-count analysis is only $\mathcal{O}(1)$ off compared to the full result.

As noted in the main text, at large signal rates (corresponding to small exposures, and hence small expected background events), the finite-background LLR slightly outperforms the zero-background projection. This is a known effect\footnote{We thank Jelle Aalbers for introducing us to this effect, as well as helping us find relevant discussions in the literature. See also their (to be published) work on the matter.}, related to the discrete nature of the median expected number of events in the Poisson distribution. At small expected background events (less than $\log2$), the median number of events is zero for the background-only hypothesis, while (possibly) increasing the $10\%$ quantile for the number of events in the background+signal hypothesis (compared to the background free case). See for example how in Ref.~\cite{Feldman:1997qc} the median limit for finite background can go below the background-free one. Broadly speaking, the discrete nature of the Poisson distribution has several unexpected features at low background expectation values, with multiple possible alternate test statistics resolving some or all of them, such as relying on the CLs method~\cite{PDG}, or utilizing Asimov data (see  Ref.~\cite{Bhattiprolu:2020mwi} which also discusses the broader issue at hand and shows the sharp features in the median estimator). However, since this point is unrelated to the main discussion of this work, we chose to use the standard sensitivity projection method reccomended in Ref.~\cite{Baxter:2021pqo}.

We will now discuss the discovery. Generally, the naive procedure of (ii) is quite close to (i), though at very large rates, it is slightly overestimating the required exposure\footnote{When less than 10 tonne-years are predicted for discovery for the NR only case, we used $10^5$ simulations, and used $Q_{2.9\times 10^{-7}}(\lambda|_{b,{\rm exp}_{\rm disc}}) \approx Q_{50\%}(\lambda|_{b,{\rm exp}_{\rm disc}})-1.6*(Q_{50\%}(\lambda|_{b,{\rm exp}_{\rm disc}})-Q_{0.1\%}(\lambda|_{b,{\rm exp}_{\rm disc}}))$ instead of our usual estimate which relies on $Q_{5\%}(\lambda|_{b,{\rm exp}_{\rm disc}}))$, because as mentioned in the main text, that estimate fails when the background rates are too low.}. The background free approximation is never valid, and (iv) is once again slightly overestimating the required exposure.

We again emphasize that none of these scenarios accounted for systematic uncertainties in the NEST code, or in the NR neutrino rate, both of which are likely to be crucial at lower signal rates.

\section{Details of the signal rate}\label{app:theory}

\subsection{Composition of xenon experiments}\label{app:composition}

Here we report the abundance of each xenon isotope in the detector, together with the corresponding mass in GeV and atomic mass number.

\begin{table}[htp!]
    \centering
    \begin{tabular}{c|c|c}
        Mass [GeV] & Atomic Mass & Fraction \\
        \hline
        119.23 & 128 & 0.019\\
        120.16 & 129 &0.264 \\
        121.09 & 130 & 0.041\\
        122.03 & 131 & 0.212\\
        122.96 & 132 & 0.269\\
        124.82 & 134 & 0.104\\
        126.68 & 136 & 0.089
    \end{tabular}
    \caption{Abundance of xenon isotopes assumed in xenon experiments.}
    \label{tab:my_label}
\end{table}

\subsection{On the discrepancies among lattice computations}\label{app:latticeerror}

As discussed in Sec.~\ref{sec:theory_err}, the present discrepancy between the central values of $\sigma_{\pi N}$ in the $N_f = 2+1+1$ and $N_f = 2+1$ flavour schemes in Eqs.\,(\ref{FLAG21}-\ref{FLAG211}) has a non-negligible impact on the total relative uncertainty on the WIMP signal rate, as clear from Fig.\,\ref{fig:deltaR}. We focus here on $\sigma_{\pi N}$ because it is the matrix element with the most significant discrepancy but similar considerations holds for $\sigma_s$. For completeness, all the lattice computations of both $\sigma_{\pi N}$ and $\sigma_s$ satisfying FLAG quality criteria are listed in Table\,\ref{lattice_sigma}. These are the results which bring to the FLAG averages in Eqs.\,(\ref{FLAG21}-\ref{FLAG211}). 

In general, $\sigma_{\pi N}$  is computed on the lattice in two ways:
\begin{itemize}
\item through the explicit computation of the matrix elements of the scalar operator as in Eq.\,\eqref{eq:hadmat}
\item via the Feynman-Hellman theorem \cite{Hellmann1937, PhysRev.56.340}, which in this context gives
\begin{equation}
\label{GMOR_eq}
\sigma_{\pi N} = m_u \frac{\partial m_N}{\partial m_u} + m_d \frac{\partial m_N}{\partial m_d} \thickapprox m_{\pi}^2 \frac{\partial m_N}{\partial m_{\pi}^2},
\end{equation}
where the last term of the equation is a consequence of the Gell-Mann--Oakes--Renner relations \cite{Gell-Mann:1968hlm}.
\end{itemize}
While the first procedure presents numerical challenges, the second one is simpler, although its result depends on the functional form assumed for the nucleon mass as a function of the pion mass. The latter is computed withing the baryonic chiral perturbation theory ($\chi$PT) whose perturbative convergence is however quite slow, resulting in large uncertainties in the functional shape of $m_N(m_\pi)$. 

All in all, we identify two interconnected effects which impact both the mean value of $\sigma_{\pi N}$ and its uncertainty in the FLAG average: $i)$ the range of the pion masses simulated on the lattice, and $ii)$ the functional structure assumed for $m_N(m_\pi)$. Regarding the pion mass, its simulated range of values influences the chiral extrapolation of lattice results to the physical point in the Feynman-Hellman approach of Eq.\,(\ref{GMOR_eq}). Refs.\,\cite{Durr:2015dna, Yang:2015uis} have considered pion masses down to 120 MeV and 139 MeV, respectively, and thus do not need any large extrapolation to the physical value of the pion mass. Conversely, in Refs.\,\cite{BMW:2011sbi, Alexandrou:2014sha} the lowest simulated pion mass value is around 200 MeV and an extrapolation is required. The extrapolation depends unavoidably on the functional shape of $m_N(m_\pi)$ which is in turn affected by the truncation order of the $\chi$PT expansions. For example Ref.\,\cite{Alexandrou:2014sha} comments explicitly on the large difference between the result obtained using $\chi$PT at order $\mathcal{O}(p^3)$ or $\mathcal{O}(p^4)$. The large systematic error associated to this lattice result try to take this discrepancy into account. 

For the reasons above we believe that further lattice studies in the $N_f = 2+1+1$ flavor scheme with pion masses closer to the physical value can shed new light on the present theoretical discrepancy with the $N_f = 2+1$ results.

\begin{table*}[t]
\begin{centering}
\renewcommand{\arraystretch}{1.2}
\begin{tabular}{|c|c|c|c||c|c|c||}
\cline{2-7}
\multicolumn{1}{c|}{} & \multicolumn{3}{c||}{$N_f = 2+1+1$}& \multicolumn{3}{c||}{$N_f = 2+1$}\\
\hline
\multicolumn{1}{|c||}{Method} & Ref. &  $\sigma_{\pi N}$& $\sigma_{s}$ & Ref. & $\sigma_{\pi N}$& $\sigma_{s}$ \\
\cline{1-7}
\multicolumn{7}{c}{}\\[-9pt]
\hline
\multicolumn{1}{|c||}{MEC} &\cite{Freeman:2012ry} & --- &0.44(8)(5) $\times m_s$ & \cite{Yang:2015uis} &  $45.9(7.4)(2.8)$ & $40.2(11.7)(3.5)$\\
\multicolumn{1}{|c||}{} &&& &\cite{Freeman:2012ry} & --- &0.637(55)(74) $\times m_s$ \\
\hline
\multicolumn{1}{|c||}{FHA} & \cite{Alexandrou:2014sha} & 64.9(1.5)(13.2) & --- & \cite{Durr:2015dna} & 38(3)(3) & 105(41)(37) \\
\multicolumn{1}{|c||}{} &&&& \cite{Junnarkar:2013ac} & --- & 48(10)(15) \\
\multicolumn{1}{|c||}{} &&&& \cite{BMW:2011sbi} & $39(4)(^{18}_7)$ & $67(27)(^{55}_{47})$ \\
\hline
\end{tabular}
\end{centering}
\caption{Overview of the lattice results for the sigma terms $\sigma_{\pi N}$ and $\sigma_{s}$ (in MeV) computed through the matrix element computation (MEC) or through the Feynman-Hellmann approach (FHA), as explained in the text. These values are the ones which pass FLAG quality criteria and which have been averaged by FLAG Collaboration separately for the cases $N_f=2+1+1$ and $N_f=2+1$. The computations in Ref.\,\cite{Freeman:2012ry} are done in the $\overline{\rm MS}$ scheme at the scale $\mu = 2$ GeV.}
\label{lattice_sigma} 
\end{table*}

\subsection{Running and matching of scalar coefficients}\label{app:running}

Here we summarize the running procedure employed to run the scalar Wilson coefficients $f_q$ and $f_G$ matched at the EW scale $\mu_Z\sim m_Z$ to the hadronic scale $\mu_{\rm had}=1$ GeV up to N$^3$LO order in $\alpha_s$. In addition to the running, threshold corrections must consistently be included at the charm and bottom quark thresholds, $\mu_c$ and $\mu_b$, respectively. This Section closely follows Ref.~\cite{Hill:2014yxa}. 

Given the scalar operators $\mathcal{O}_q$ and $\mathcal{O}_G$, the equivalence of matrix elements computed at different scales $\mu_h>\mu_l$ or with different numbers of active quarks enforces the following running and matching relations:
\begin{equation}
\begin{split}
\label{runmatcon}
&\langle \mathcal{O}_i\rangle(\mu_h)=R_{ji}(\mu_l,\mu_h)\langle \mathcal{O}_j\rangle(\mu_l)\,, \\
&\langle \mathcal{O}_i'\rangle(\mu_Q) = M_{ji}(\mu_Q)\langle \mathcal{O}_j\rangle(\mu_Q)\ ,
\end{split}
\end{equation}
where $i,\,j=q,\,G$ and $\mu_Q=\mu_c,\mu_b$ is a heavy quark threshold. The primed operator refers to $n_f+1$ flavors, while the unprimed one to $n_f$ flavors, while $\langle\mathcal{O}\rangle\equiv \langle N|\mathcal{O}|N\rangle$. In terms of the Wilson coefficients, the relations above are equivalent to
\begin{equation}
\begin{split}
&f_i(\mu_l)=R_{ij}(\mu_l,\mu_h)f_j(\mu_h),\, \\
&f_i(\mu_q) = M_{ij}(\mu_q)f_j'(\mu_q)\, .
\end{split}
\end{equation}

The running of the matrix elements of the scalar operators are simplified by few non-perturbative relations. First,  $\mathcal{O}_q$ is scale invariant so the running matrix can be immediately written as

\begin{equation}
R(\mu_l,\mu_h)=
\begin{pmatrix}
  \begin{matrix}
  1 & &  \\
    & \ddots & \\
    & & 1
  \end{matrix}
  & \rvline & \begin{matrix}
  R_{qG}\\
  \vdots\\
  R_{qG}
  \end{matrix} \\
\hline
  \begin{matrix}
  0&\cdots &0
\end{matrix}   & \rvline &
  R_{GG}
\end{pmatrix}\ .
\end{equation}

Second, the scale invariance of the trace of the energy-momentum tensor of QCD implies that

\begin{equation}
\langle\Theta_\mu^\mu\rangle=m_N=\frac{\pi\tilde{\beta}(g_s)}{2\alpha_s}\langle O_G\rangle+(1-\gamma_m)\sum_q\langle O_q\rangle\ ,
\end{equation}
where $\tilde{\beta}(g_s) \equiv \beta(g_s) / g_s$, $\beta(g_s)$ is the QCD beta function, and $\gamma_m$ is the quark mass renormalization function. All in all we get 
\begin{equation*}
\begin{split}
&R_{GG}=\frac{\tilde{\beta}(\mu_l)}{\tilde{\beta}(\mu_h)}\frac{\alpha_s(\mu_h)}{\alpha_s(\mu_l)}\,, \\
&R_{qG}=\frac{2\alpha_s(\mu_h)}{\pi\tilde{\beta}(\mu_h)}(\gamma_m(\mu_h)-\gamma_m(\mu_l))\ .
\end{split}
\end{equation*}

Concerning the mathcing functions $M_{ij}$, the chiral limit on the light quarks $q,~q'$ requires $M_{qq'}=M_q\delta_{qq'}+O(1/m_Q)$, $Q$ being  the heavy quark. Moreover, the quark masses are defined to include the effects coming from the heavy quark $Q$ (that is integrated out), so that we can fix $M_q=1$. Thus, from Eq.~\eqref{runmatcon} we get
\begin{equation}
\langle  \mathcal{O}_q'\rangle=M_{Gq}\langle  \mathcal{O}_G\rangle+\langle  \mathcal{O}_q\rangle\,.
\end{equation}
Dimensional analysis suggests that
\begin{equation}
M_{Gq}\sim \left.\frac{\langle \mathcal{O}_q\rangle}{\langle \mathcal{O}_G\rangle}\right|_{m_Q}\propto \frac{m_q}{m_Q}\ ,
\end{equation}
so that we can neglect $M_{Gq}$ and we can write explicitly the matrix $M$ as
\begin{equation}
M(\mu_Q)=
\begin{pmatrix}
  \begin{matrix}
  1 & &  \\
    & \ddots & \\
    & & 1
  \end{matrix}
  & \rvline & \begin{matrix}
  M_{qQ}\\
  \vdots\\
  M_{qQ}
  \end{matrix} & \rvline & \begin{matrix}
  M_{qG}\\
  \vdots\\
  M_{qG}
  \end{matrix}\\
\hline
  \begin{matrix}
  0&\cdots &0
\end{matrix}   & \rvline &
  M_{GQ} & \rvline &
  M_{GG}
\end{pmatrix}\ .
\end{equation}
 Finally, two more relations come from the scale invariance of the energy momentum tensor and read

\begin{align}
&(1-\gamma_m^{(n_f+1)})(1+M_{qQ})+\frac{\pi\tilde{\beta}^{(n_f+1)}}{2\alpha_s^{(n_f+1)}}M_{qG}=1-\gamma_m^{(n_f)}\, ,\notag\\
&(1-\gamma_m^{(n_f+1)})M_{GQ}+\frac{\pi\tilde{\beta}^{(n_f+1)}}{2\alpha_s^{(n_f+1)}}M_{GG}=\frac{\pi\tilde{\beta}^{(n_f)}}{2\alpha_s^{(n_f)}}\, , \label{eqdiffM}
\end{align}
where we have explicitly shown as superscripts the number of dynamical flavors used to evaluate the beta function, the anomalous dimensions and the coupling. From Eq.~\eqref{eqdiffM}, we can compute $M_{GG}$ and $M_{qG}$ in terms of $M_{qQ}$ and $M_{GQ}$, whose expressions were found up to N$^3$LO in Ref.~\cite{Chetyrkin:1997un}. Upon expanding all the matching and running coefficients to the same order, the relation between the Wilson coefficients at the hadronic and at the EW scale is given by

\begin{align}
  f_j(\mu_{\rm had}) &= R_{jk}(\mu_{\rm had}, \mu_c)M_{kl}(\mu_c)\times\label{eq:wilsonrun}\\
  &\times R_{lm}(\mu_c,\mu_b)M_{mn}(\mu_b)R_{ni}(\mu_b, \mu_Z)f_i(\mu_Z) \ .\notag
\end{align}

\subsection{Uncertainties on the SI cross-section}\label{sec:data}

The theoretical uncertainty on the SI cross-section is determined by: $i)$ the truncation of the perturbative expansion at N$^3$LO in $\alpha_s$ in the scalar amplitude, $ii)$ the lattice uncertainties on the sigma terms in Eq.~\eqref{FLAG21} and Eq.~\eqref{FLAG211} affecting again the scalar amplitude, $iii)$ the experimental uncertainties on the second moments of the PDFs of quark, antiquark and gluon, as reported in Table~\ref{T2unc}, which affect the twist-2 amplitude.

We estimate the uncertainty $i)$ by sampling the three relevant matching scales in Eq.~\eqref{eq:wilsonrun} with a uniform distribution within the intervals
\begin{equation}
\label{varmatch}
\begin{split}
&\frac{m_Z}{2}<\mu_Z<m_Z\, ,\\ 
&2m_c<\mu_b<2m_b\, ,\\
&\mu_{\rm had}<\mu_c<2m_c\ ,
\end{split}
\end{equation}
where $\mu_{\rm had} = 1\,\text{GeV}$ and $m_Z=91.1876\text{ GeV}$, $m_b=4.18\text{ GeV}$, $m_c=1.27\text{ GeV}$.
These values are taken from the PDG \cite{PDG}, in particular the $b$- and the $c$-quark masses are determined in the $\overline{\rm MS}$ scheme. 

Similarly, since the uncertainties in Table \ref{T2unc} and in Eqs.\,(\ref{FLAG21}-\ref{FLAG211}) are Gaussian, we determine the impact of the uncertainty on the second moments of the PDFs and the sigma terms by sampling  them through appropriate Gaussian distributions. The resulting distribution of the SI cross-section from the simultaneous sampling of all the aformentioned parameters was observed to be Gaussian for all the EW multiplets except for the $2_{1/2}$, because of the large accidental cancellations discussed in Section \ref{sec:theory_err}.\\

In order to estimate the relative importance of the different sources of uncertainty, we also sampled each group of parameters separately while keeping the others fixed to their central values. We summarize all these results in Tables \ref{tab:211} and \ref{tab:21} for $N_f=2+1+1$ and $N_f=2+1$ lattice schemes, respectively, where the single uncertainty from running, lattice and PDFs are denoted as $\Delta \sigma_\mu$, $\Delta \sigma_{f_T}$, $\Delta \sigma_{\rm T2}$, respectively, while the total uncertainty as $\Delta\sigma_N$. As we can see, while in the $N_f=2+1$ case all the uncertainties are roughly of the same order, the lattice error generally dominates in the $N_f=2+1+1$ case due to the large systematic uncertainty in $\sigma_{\pi N}$ included by \cite{Alexandrou:2014sha}, as discussed in Section \ref{app:latticeerror}.

\begin{table}[t]
\caption{Second moments of the PDFs of quark, antiquark and gluon at the EW scale used in this work.}\label{tab:PDFmoments}
\renewcommand{\arraystretch}{1.3}
\centering
\begin{tabular}{c||c} 
\hline
$g(2,m_Z)=0.464\pm 0.002$ &  \\
$u(2,m_Z)=0.223\pm 0.003$ & $\overline{u}(2,m_Z)=0.036\pm 0.002$\\
$d(2,m_Z)=0.118\pm 0.003$ & $\overline{d}(2,m_Z)=0.037\pm 0.003$ \\
$s(2,m_Z)=0.0258\pm 0.0004$ & $\overline{s}(2,m_Z)=0.0258\pm 0.0004$ \\
$c(2,m_Z)=0.0187\pm 0.0002$ & $\overline{c}(2,m_Z)=0.0187\pm 0.0002$ \\
$b(2,m_Z)=0.0117\pm 0.0001$ & $\overline{b}(2,m_Z)=0.0117\pm 0.0001$  \\
\hline
\end{tabular} 
\label{T2unc}
\end{table}

\begin{table*}[t]
\begin{centering}
\renewcommand{\arraystretch}{1.2}
\begin{tabular}{|c|c|c|c|c|c|}
\cline{2-6}
\multicolumn{1}{c|}{} & \multicolumn{5}{c|}{$N_f = 2+1+1$}\\
\hline
\multicolumn{1}{|c||}{$n$} & $\sigma_N$ & $\Delta \sigma_\mu$& $\Delta \sigma_{f_T}$& $\Delta \sigma_{\rm T2}$&$\Delta \sigma_N$ \\
\cline{1-6}
\multicolumn{6}{c}{}\\[-9pt]
\hline
\multicolumn{1}{|c||}{$2_{1/2}$} & $- $ &  $<0.31\times 10^{-50}$& $<1.9\times 10^{-50}$& $<0.5\times 10^{-50}$ & $2.3 \times 10^{-50}$ \\
\cline{2-6}
\multicolumn{1}{|c||}{3$_0$}    & $1.87\times 10^{-47}$ &  $0.12\times 10^{-47}$& $0.22\times 10^{-47}$ & $0.18\times 10^{-47}$ & $0.3\times 10^{-47}$\\
\cline{2-6}
\multicolumn{1}{|c||}{$3_{1}$}  & $5.0\times 10^{-48}$  &  $0.4\times 10^{-48}$ & $1.8\times 10^{-48}$  & $0.8\times 10^{-48}$  & $2.0\times 10^{-48}$ \\
\cline{2-6}
\multicolumn{1}{|c||}{$4_{1/2}$}& $4.2\times 10^{-47}$  &  $0.3\times 10^{-47}$ & $0.7\times 10^{-47}$  & $0.5\times 10^{-47}$  & $1.0\times 10^{-47}$ \\
\cline{2-6}
\multicolumn{1}{|c||}{5$_0$}    & $1.69\times 10^{-46}$ &  $0.11\times 10^{-46}$& $0.20\times 10^{-46}$ & $0.16\times 10^{-46}$ & $0.28\times 10^{-46}$\\
\cline{2-6}
\multicolumn{1}{|c||}{$5_{1}$}  & $5.1\times 10^{-47}$  &  $0.5\times 10^{-47}$ & $1.2\times 10^{-47}$  & $0.8\times 10^{-47}$  & $1.6\times 10^{-47}$ \\
\cline{2-6}
\multicolumn{1}{|c||}{$6_{1/2}$}& $3.0\times 10^{-46}$  &  $0.2\times 10^{-46}$ & $0.4\times 10^{-46}$  & $0.3\times 10^{-46}$  & $0.6\times 10^{-46}$ \\
\cline{2-6}
\multicolumn{1}{|c||}{7$_0$}    & $6.8\times 10^{-46}$  &  $0.4\times 10^{-46}$ & $0.8\times 10^{-46}$  & $0.6\times 10^{-46}$  & $1.1\times 10^{-46}$ \\
\cline{2-6}
\multicolumn{1}{|c||}{$8_{1/2}$}& $1.05\times 10^{-45}$ &  $0.07\times 10^{-45}$& $0.14\times 10^{-45}$ & $0.10\times 10^{-45}$ & $0.17\times 10^{-45}$\\
\cline{2-6}
\multicolumn{1}{|c||}{9$_0$}    & $1.88\times 10^{-45}$ &  $0.12\times 10^{-45}$&$0.24\times 10^{-45}$  & $0.18\times 10^{-45}$ & $0.3\times 10^{-45}$\\
\cline{2-6}
\multicolumn{1}{|c||}{10$_{1/2}$}& $2.7\times 10^{-45}$ &  $0.2\times 10^{-45}$ & $0.3\times 10^{-45}$  & $0.3\times 10^{-45}$  & $0.4\times 10^{-45}$ \\
\cline{2-6}
\multicolumn{1}{|c||}{11$_0$}    & $4.2\times 10^{-45}$ &  $0.3\times 10^{-45}$ & $0.5\times 10^{-45}$  & $0.4\times 10^{-45}$  & $0.7\times 10^{-45}$ \\
\cline{2-6}
\multicolumn{1}{|c||}{$12_{1/2}$}& $5.7\times 10^{-45}$ &  $0.4\times 10^{-45}$ & $0.7\times 10^{-45}$  & $0.6\times 10^{-45}$  & $0.9\times 10^{-45}$ \\
\cline{2-6}
\multicolumn{1}{|c||}{13$_0$}    & $8.3\times 10^{-45}$ &  $0.5\times 10^{-45}$ & $1.0\times 10^{-45}$  & $0.9\times 10^{-45}$  & $1.4\times 10^{-45}$ \\
\hline
\end{tabular}
\end{centering}
\caption{Mean values and uncertainties of the SI elastic cross-sections (in cm$^2$) for each $n$-plet, adopting the $N_f$=2+1+1  values of the sigma terms in Eq.~\eqref{FLAG211}. }
\label{tab:211} 
\end{table*}

\begin{table*}[t]
\begin{centering}
\renewcommand{\arraystretch}{1.2}
\begin{tabular}{|c|c|c|c|c|c|}
\cline{2-6}
\multicolumn{1}{c|}{} & \multicolumn{5}{c|}{$N_f = 2+1$}\\
\hline
\multicolumn{1}{|c||}{$n$} & $\sigma_N$ & $\Delta \sigma_\mu$& $\Delta \sigma_{f_T}$& $\Delta \sigma_{\rm T2}$&$\Delta \sigma_N$ \\
\cline{1-6}
\multicolumn{6}{c}{}\\[-9pt]
\hline
\multicolumn{1}{|c||}{$2_{1/2}$} & $1.8 \times 10^{-50}$ &  $^{+1.2}_{-0.8} \times 10^{-50}$& $<2.6\times 10^{-50}$& $<2.6\times 10^{-50}$& $<2.8\times 10^{-50}$ \\
\cline{2-6}
\multicolumn{1}{|c||}{3$_0$}     & $2.06\times 10^{-47}$ &  $0.12\times 10^{-47}$  & $0.12\times 10^{-47}$  & $0.18\times 10^{-47}$& $0.24\times 10^{-47}$ \\
\cline{2-6}
\multicolumn{1}{|c||}{$3_{1}$}   & $3.5\times 10^{-48}$ &  $0.3\times 10^{-48}$& $0.7\times 10^{-48}$& $0.6\times 10^{-48}$& $1.0\times 10^{-48}$ \\
\cline{2-6}
\multicolumn{1}{|c||}{$4_{1/2}$}  & $4.8\times 10^{-47}$ &  $0.3\times 10^{-47}$&$0.3\times 10^{-47}$& $0.5\times 10^{-47}$& $0.7\times 10^{-47}$ \\
\cline{2-6}
\multicolumn{1}{|c||}{5$_0$}  &  $1.86\times 10^{-46}$ &  $0.11\times 10^{-46}$& $0.11\times 10^{-46}$& $0.18\times 10^{-46}$& $0.22\times 10^{-46}$ \\
\cline{2-6}
\multicolumn{1}{|c||}{$5_{1}$}  & $6.3\times 10^{-47}$ &  $0.5\times 10^{-47}$& $0.7\times 10^{-47}$& $0.9\times 10^{-47}$& $1.3\times 10^{-47}$ \\
\cline{2-6}
\multicolumn{1}{|c||}{$6_{1/2}$}  & $3.4\times 10^{-46}$ &  $0.2\times 10^{-46}$& $0.2\times 10^{-46}$& $0.4\times 10^{-46}$& $0.4\times 10^{-46}$ \\
\cline{2-6}
\multicolumn{1}{|c||}{7$_0$}  & $7.5\times 10^{-46}$ &  $0.4\times 10^{-46}$&$0.7\times 10^{-46}$& $0.9\times 10^{-46}$& $0.9\times 10^{-46}$ \\
\cline{2-6}
\multicolumn{1}{|c||}{$8_{1/2}$}  &  $1.17\times 10^{-45}$ &  $0.07\times 10^{-45}$& $0.08\times 10^{-45}$& $0.11\times 10^{-45}$& $0.13\times 10^{-45}$ \\
\cline{2-6}
\multicolumn{1}{|c||}{9$_0$}  &  $2.07\times 10^{-45}$ &  $0.12\times 10^{-45}$& $0.12\times 10^{-45}$& $0.18\times 10^{-45}$& $0.22\times 10^{-45}$ \\
\cline{2-6}
\multicolumn{1}{|c||}{$10_{1/2}$}  & $3.0\times 10^{-45}$ &  $0.2\times 10^{-45}$&$0.2\times 10^{-45}$& $0.3\times 10^{-45}$& $0.3\times 10^{-45}$ \\
\cline{2-6}
\multicolumn{1}{|c||}{11$_0$}  &  $4.7\times 10^{-45}$ &  $0.3\times 10^{-45}$&$0.3\times 10^{-45}$& $0.5\times 10^{-45}$& $0.6\times 10^{-45}$ \\
\cline{2-6}
\multicolumn{1}{|c||}{$12_{1/2}$}  & $6.4\times 10^{-45}$ &  $0.4\times 10^{-45}$& $0.4\times 10^{-45}$& $0.6\times 10^{-45}$& $0.8\times 10^{-45}$ \\
\cline{2-6}
\multicolumn{1}{|c||}{13$_0$}  & $9.2\times 10^{-45}$ &  $0.5\times 10^{-45}$& $0.5\times 10^{-45}$& $0.8\times 10^{-45}$& $1.1\times 10^{-45}$ \\
\hline
\end{tabular}
\end{centering}
\caption{Mean values and uncertainties of the SI elastic cross-sections (in cm$^2$) for each $n$-plet, adopting the  $N_f$=2+1 values of the sigma terms in Eq.~\eqref{FLAG21}. }
\label{tab:21} 
\end{table*}

\subsection{Wilson Coefficients at the EW scale}
\label{WClimit}

We collect here the explicit expressions of the coefficients $f_q$, $f_G$, $g_{1,2}^{q}$ and $g_{1,2}^{G}$, which acquire simple forms in the limit $M_\chi\gg m_W,\,m_Z,\,m_t$. These results were first computed in Ref.~\cite{Hisano:2015rsa}.
The quark coefficients read:
\begin{equation*}
\begin{split}
f_q(m_Z)&\simeq-\frac{\pi\alpha_2^2(n^2-1-4 Y^2)}{16\,m_h^2\,m_W}\left(1+ 0.39 \alpha_s \right)\, \\
&\hskip -0.9truecm + \frac{\pi \alpha_2^2Y^2}{4 c_w^4 m_Z^3}[-0.53+a_{Vq}^2(4+0.42\alpha_s)-a_{Aq}^2(4+2.97 \alpha_s)] ,
\end{split}
\end{equation*}
\begin{equation*}
\begin{split}
f_b(m_Z)&=-\frac{\pi\alpha_2^2(n^2-1-4 Y^2)}{16\,m_h^2\,m_W}\left(1+ 0.003 \alpha_s \right)\, \\
&\hskip -0.9truecm + \frac{\pi \alpha_2^2Y^2}{4 c_w^4 m_Z^3}[-0.53+a_{Vb}^2(4+0.42\alpha_s)-a_{Ab}^2(4+2.97 \alpha_s)] ,
\end{split}
\end{equation*}
\begin{equation*}
\begin{split}
g_1^q(m_Z)&\simeq\frac{\alpha_2^2(n^2-1-4 Y^2)}{8 m_W^3}\left(1.05+ 0.85 \alpha_s \right)\, \\
&\hskip -0.9truecm + \frac{\alpha_2^2Y^2}{c_w^4 m_Z^3}[a_{Vq}^2(4.19+3.12\alpha_s)+a_{Aq}^2(4.19+3.12 \alpha_s)] ,
\end{split}
\end{equation*}
\begin{equation*}
\begin{split}
g_1^b(m_Z)&\simeq\frac{\alpha_2^2(n^2-1-4 Y^2)}{8 m_W^3}\left(0.14 - 0.005 \alpha_s \right)\, \\
&\hskip -0.9truecm + \frac{\alpha_2^2Y^2}{c_w^4 m_Z^3}[a_{Vb}^2(4.19+3.12\alpha_s)+a_{Ab}^2(4.19+3.12 \alpha_s)] ,
\end{split}
\end{equation*}
\begin{equation*}
\begin{split}
g_2^q(m_Z)&\simeq\frac{\alpha_2^2(n^2-1-4 Y^2)}{108 m_W^3}\left(0+ \alpha_s \right)\, \\
&\hskip -0.9truecm + \frac{\alpha_2^2Y^2}{c_w^4 m_Z^3}[0.297 a_{Vq}^2 (0+\alpha_s)+0.297 a_{Aq}^2 (0+\alpha_s)] ,
\end{split}
\end{equation*}
\begin{equation*}
\begin{split}
g_2^b(m_Z)&\simeq \frac{\alpha_2^2Y^2}{c_w^4 m_Z^3}[0.297 a_{Vb}^2 (0+\alpha_s)+0.297 a_{Ab}^2 (0+\alpha_s)] ,
\end{split}
\end{equation*}
where in the above expressions $q = u,\,d,\,s,\,c$. Here $a_{Vq} = T_{3q}/2 - Q_q s_w^2$, $a_{Aq} =-T_{3q}/2$, $a_{Vb} = -1/4 + s_w^2/3$, $a_{Ab} =1/4$ with $c_w,\,s_w$ being the cosine and the sine of the Weinberg angle, respectively.

The gluon coefficients, instead, are:
\begin{equation*}
\begin{split}
f_G(m_Z)&\simeq\frac{\alpha_2^2(n^2-1-4 Y^2)}{32\,m_h^2\,m_W}\left(1.86 + 0.93 \alpha_s \right)\, \\
&\hskip -0.9truecm + \frac{\alpha_2^2Y^2}{8 c_w^4 m_Z^3}[0.28(1+0.88\alpha_s)-1.06 a_{Vt}^2+3.09 a_{At}^2\, \\
&\hskip -0.9truecm +1.05\left(\sum_{q} (a_{Vq}^2+a_{Aq}^2)\right)(1+0.37\alpha_s)] ,
\end{split}
\end{equation*}
\begin{equation*}
\begin{split}
g_1^G(m_Z)&\simeq\frac{\alpha_2^2(n^2-1-4 Y^2)}{32 m_W^3}\left(0+ 2.66 \alpha_s \right)\, \\
&\hskip -0.9truecm + \frac{\alpha_2^2Y^2}{4 c_w^4 m_Z^3}[\left(\sum_{q} (a_{Vq}^2+a_{Aq}^2)\right)(0+ 2.44 \alpha_s )] ,
\end{split}
\end{equation*}
\begin{equation*}
\begin{split}
g_2^G(m_Z)&\simeq-\frac{\alpha_2^2(n^2-1-4 Y^2)}{72 m_W^3}\left(0+ \alpha_s \right)\, \\
&\hskip -0.9truecm - \frac{\alpha_2^2Y^2}{9 c_w^4 m_Z^3}[\left(\sum_{q} (a_{Vq}^2+a_{Aq}^2)\right)(0+ \alpha_s )] ,
\end{split}
\end{equation*}
where the sum now runs over all the active quarks, namely $q = u,\,d,\,s,\,c,\,b$ and $a_{Vt} = 1/4 - 2 s_w^2/3$, $a_{At} =-1/4$.

\end{document}